\documentclass[12pt,a4]{article}
\usepackage{amsthm,amsmath,latexsym,amssymb,amsfonts,amscd}
\usepackage{graphics,lscape,fancyhdr,array,stmaryrd,euscript}
\usepackage{empheq}
\usepackage{sidecap}
\usepackage{ifpdf}
\usepackage{verbatim}
\usepackage{color}
\usepackage{relsize}
\usepackage{tikz-cd}
\numberwithin{equation}{section}
\usepackage{hyperref,setspace}

\usepackage[numbers,sort&compress]{natbib}
\usepackage[nottoc]{tocbibind}

\newcommand{\pl}{\partial}

\newcommand{\be}{\begin{equation}}
\newcommand{\ee}{\end{equation}}

\newcommand{\besubeqs}{\begin{subequations}}
\newcommand{\esubeqs}{\end{subequations}}

\newcommand{\zb}{{\bar{z}}}

\newcommand{\pb}{{\bar{p}}}
\newcommand{\qb}{{\bar{q}}}
\newcommand{\Jb}{{\bar{J}}}
\newcommand{\jb}{{\bar{j}}}

\newcommand{\plb}{{\bar{\pl}}}

\newcommand{\pfrac}[1]{{\frac{\pl}{\pl #1}}}

\newcommand{\deltas}[1]{{\delta\left(#1\right)}}
\newcommand{\PP}{{\mathbb{P}}}

\newcommand{\PPb}{{\bar{\mathbb{P}}}}

\newcommand{\JJJ}{{\bold{J}}}
\newcommand{\HHH}{{\bold{H}}}

\newcommand{\inter}{_{\text{int}}}

\begin{document}
\thispagestyle{empty}
\hfill
\begin{flushright}
    {LMU-ASC 45/16}\\
    {Imperial-TP-DP-2016-01}
\end{flushright}
\vskip 0.015\textheight
\begin{center}

{\Large\bfseries Light-Front Higher-Spin Theories in Flat Space} \\
\vskip 0.03\textheight

Dmitry \textsc{Ponomarev},${}^{1}$ Evgeny \textsc{Skvortsov},${}^{2,3}$

\vskip 0.03\textheight

{\em ${}^{1}$Theoretical physics group, Blackett Laboratory,\\ Imperial College London, SW7 2AZ, U.K.}\\
\vspace*{5pt}
{\em ${}^{2}$ Arnold Sommerfeld Center for Theoretical Physics\\
Ludwig-Maximilians University Munich\\
Theresienstr. 37, D-80333 Munich, Germany}\\

\vspace*{5pt}
{\em ${}^{3}$ Lebedev Institute of Physics, \\
Leninsky ave. 53, 119991 Moscow, Russia}

\vskip 0.02\textheight

{\bf Abstract }

\end{center}
\begin{quotation}
\noindent
We revisit the problem of interactions of higher-spin fields in flat space. We argue that all no-go theorems can be avoided by the light-cone approach, which results in more interaction vertices as compared to the usual covariant approaches. It is stressed that there exist two-derivative gravitational couplings of higher-spin fields. We show that some reincarnation of the equivalence principle still holds for higher-spin fields --- the strength of gravitational interaction does not depend on spin. Moreover, it follows from the results by Metsaev that there exists a complete chiral higher-spin theory in four dimensions. We give a simple derivation of this theory and show that the four-point scattering amplitude vanishes. Also, we reconstruct the quartic vertex of the scalar field in the unitary higher-spin theory, which turns out to be perturbatively local.
\indent 
\end{quotation}

\newpage
\clearpage
\setcounter{page}{1}
\tableofcontents

\section{Introduction}
Since the early days of quantum field theory there have been many no-go results that prevent non-trivial interacting theories with massless higher-spin fields to exist. Notable examples are the Weinberg low energy theorem \cite{Weinberg:1964ew} and the Coleman-Mandula theorem \cite{Coleman:1967ad}. One possible way out is to switch on the cosmological constant \cite{Fradkin:1986qy,Vasiliev:1990en,Vasiliev:2003ev}, which simultaneously avoids the no-go theorems that are formulated for QFT in flat space. Higher-spin theories in anti-de Sitter space later received a solid ground on the base of AdS/CFT correspondence \cite{Maldacena:1997re,Gubser:1998bc,Witten:1998qj} where higher-spin theories are supposed to be generic duals of free CFT's \cite{Klebanov:2002ja,Sezgin:2002rt,Sezgin:2003pt,Giombi:2009wh} with certain interacting ones accessible via an alternate choice \cite{Klebanov:1999tb} of boundary conditions \cite{Klebanov:2002ja,Sezgin:2003pt,Leigh:2003gk,Giombi:2009wh,Giombi:2011ya}. 

The fate of higher-spin theories in flat space is still unclear and is a source of controversy. The no-go theorems are still true. Also, within the local field theory approach one immediately faces certain obstructions: Aragone-Deser argument forbids minimal gravitational interactions of massless higher-spin fields \cite{Aragone:1979hx,Aragone:1981yn} and, even if relaxing this assumption, it is still impossible to deform the gauge algebra \cite{Bekaert:2010hp,Joung:2013nma}. These results are based on the gauge invariant and manifestly Lorentz covariant field description in terms of Fronsdal fields \cite{Fronsdal:1978rb}, which suggests another possible way out.

Indeed, gauge symmetry can be thought of as just a redundancy of description, though it turns out to be  exceptionally useful in many cases. Therefore, in order to look for higher-spin theories in flat space it can be useful to turn to methods that deal with physical degrees of freedom only and thereby avoid any problems that originate from specific field descriptions. One such method is the light-cone approach, which still allows one to have a local field theory.

It is in the light-cone approach that the first examples of non-trivial cubic interactions between higher-spin fields were found in \cite{Bengtsson:1983pg,Bengtsson:1983pd,Bengtsson:1986kh}. The covariant results followed soon after \cite{Berends:1984wp,Berends:1984rq}. A detailed classification of cubic vertices within the light cone approach is now available in all dimensions for massive and massless fields of arbitrary spin and symmetry type \cite{Fradkin:1991iy,Metsaev:1993ap,Metsaev:2005ar,Metsaev:2007rn}.

In this paper we revisit the problem of constructing higher-spin theories in flat space, specifically in four-dimensions. First of all, we argue that at least formally the most powerful no-go theorems are avoided by the light-cone approach. Also, we recall that there is a mismatch between the covariant cubic vertices and those found in \cite{Bengtsson:1983pg,Bengtsson:1983pd,Bengtsson:1986kh}
by the light-cone methods: there exist exceptional vertices not seen by some of the covariant methods. In particular, there does exist a two-derivative gravitational vertex for a field of any spin  \cite{Bengtsson:2014qza, Conde:2016izb,Sleight:2016xqq}, which is also evident in the language of amplitudes \cite{Benincasa:2011pg,Benincasa:2007xk}. 

Having the gravitational higher-spin vertex at our disposal we prove that fields of any spin couple to gravity universally, i.e. some form of the equivalence principle is still true for higher-spin fields. In fact, the strength of the gravitational coupling does not depend on spin at all.

A remarkable result obtained by Metsaev in \cite{Metsaev:1991nb,Metsaev:1991mt} is that one can fix the cubic vertex without having to perform the full quartic analysis. We present a simple derivation of this result, which clarifies the assumptions. Based on this solution, we note that there exists a consistent non-trivial higher-spin theory in flat space. This theory contains graviton, massless higher-spin fields, the two-derivative gravitational vertices as well as other vertices. The action terminates at cubic vertices. Like in the self-dual Yang-Mills theory the four-point scattering amplitude vanishes. The only feature is that it breaks parity and is non-unitary. Nevertheless, it provides a counterexample to a widespread belief that higher-spin theories in flat space do not exist at all.  

Aiming at the unitary and parity preserving higher-spin theory in flat space we reconstruct the part of the quartic Hamiltonian that contains self-interactions of the scalar field, which can be regarded as the flat space counterpart of the $AdS_4$ result \cite{Bekaert:2014cea,Bekaert:2015tva}.

The outline is as follows. In Section \ref{sec:notsonogo}
we discuss how to avoid the famous no-go results. In Section \ref{sec:lightfront} we review the basics of the light-cone approach with the main result being the classification of all possible couplings that was obtained in \cite{Bengtsson:1983pg,Bengtsson:1983pd,Bengtsson:1986kh,Metsaev:1991nb,Metsaev:1991mt}. The relation to the Lorentz covariant classification is spelled out in Section \ref{subsec:covariant}. In Section \ref{sec:selfdual} we present a complete chiral higher-spin theory with the details of the derivation of the Metsaev solution \cite{Metsaev:1991nb,Metsaev:1991mt} devoted to Appendix \ref{app:insym}. The scalar part of the quartic Hamiltonian of the unitary higher-spin theory is reconstructed in Section \ref{sec:quarticH}. The higher-spin equivalence principle is derived in Section \ref{sec:HSequivalence}. We conclude with some discussion of possible extensions of these results in Section \ref{sec:conclusions}.

\section{Avoiding No-Go Theorems}
\label{sec:notsonogo}
In the distant past it was a common belief that higher-spin theories, i.e. the theories with massless fields with spin greater than two, are not consistent. The most notable examples of such no-go theorems are Weinberg low energy theorem \cite{Weinberg:1964ew}, Coleman-Mandula theorem \cite{Coleman:1967ad} and the Aragone-Deser argument \cite{Aragone:1979hx}. We briefly discuss them below, see also a very nice review \cite{Bekaert:2010hw}, as to point out how all of them can be avoided.

Our conclusion is that there are still good chances to have nontrivial higher-spin theories in flat space. Moreover, we will present an example of consistent chiral theory in Section \ref{sec:selfdual}. However, it should be stressed that while higher-spin theories may avoid the assumptions of the no-go theorems they may not defy the spirit of these theorems: there are strong indications that $S$-matrix should be trivial in some sense. For example, for the case of conformal higher-spin theories the $S$-matrix is a combination of $\delta(s,t,u)$  \cite{Beccaria:2016syk,Joung:2015eny,Sleight:2016xqq} and the AdS/CFT duals of unbroken higher-spin theories must be free CFT's \cite{Maldacena:2011jn,Alba:2013yda,Boulanger:2013zza,Stanev:2013qra,Alba:2015upa}, which should be thought of as examples of trivial holographic $S$-matrices.

\paragraph{Weinberg low energy theorem.} A serious restriction comes from the Weinberg low energy theorem \cite{Weinberg:1964ew} that eventually leads to too many conservation laws, when massless higher-spin fields are present. As a result of checking linearized gauge invariance or Lorentz invariance of the $n$-particle amplitude with one soft spin-$s$ particle attached one finds
\begin{align} &\parbox{4cm}{\includegraphics[scale=0.15]{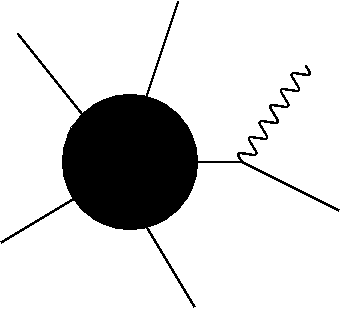}}&& \Longleftrightarrow&&\sum_i g^i_s\, p^i_{\mu_1} ...p^i_{\mu_{s-1}}=0\end{align}
where $g_s^i$ is the coupling constant of the $i$-th species to a spin-$s$ field. For $s=1$ one discovers that the total (electric) charge is conserved. For $s=2$ one finds a linear combination of momenta weighted by $g_2^i$ whose clash with the momentum conservation law $\sum_i p^i_\mu=0$ can only be resolved by the equivalence principle, i.e. all fields must couple to gravity universally, $g^i_2=const$. 

For the higher-spin case $s>2$ one finds too many conservations laws, which is a rank $(s-1)$ tensorial expression, with the only solution given by permutations of momenta at the condition that all coupling constants are the same. 

In the course of the proof of the theorem one makes an explicit use of Lorentz covariant vertices. In particular, the expressions are manifestly Lorentz covariant. This is not the case in the light-cone approach where the vertices do not have a manifestly Lorentz covariant form. It would be interesting to reconsider the Weinberg theorem as to see whether these assumptions can be weakened.\footnote{We are grateful to Sasha Zhiboedov for the useful discussion of this problem.}

\paragraph{Coleman-Mandula theorem.} The famous Coleman-Mandula theorem \cite{Coleman:1967ad} prevents $S$-matrix from having symmetry generators, beyond those of the Poincare group, that transform under the Lorentz group. Under assumptions of non-triviality of the symmetry action, discrete mass spectrum and the analyticity of the $S$-matrix in Mandelstam invariants, it can be shown that the symmetry algebra can only be a product of the Poincare group and a group of internal symmetries whose generators are Lorentz scalars. It does not apply to the case of $d=1+1$ QFT, where only forward/backward scattering is possible, so $S$-matrix must have scattering angles $\theta=0,\pi$ and thereby it is not analytic. The essence of the proof is that the scattering process is a map from one set of momenta to another one and the momenta are restricted by energy-momentum conservation, which is a Lorentz vector equation. Existence of some other charges that transform non-trivially under the spacetime symmetry would impose tensorial equations on momenta, e.g. like in Weinberg theorem, which would restrict possible processes to exchanges of momenta like in $1+1$ or trivialize the scattering completely. One way the original Coleman-Mandula theorem can be avoided is by assuming that symmetry generators transform as spinors, which leads to supersymmetry.

One of the assumptions of the theorem is to have a finite number of particles below any mass-shell. This is certainly not true in higher-spin theories where the spectrum should contain infinitely many massless particles \cite{Fronsdal:1978vb,Berends:1984rq,Fradkin:1986ka,Boulanger:2013zza}. It would be interesting to weaken the assumptions of the theorem \cite{commMassimo}. 

\paragraph{Aragone-Deser argument/No canonical gravity coupling.} Contrary to the Weinberg and Coleman-Mandula theorems, this argument is local and is attached to specific field variables \cite{Aragone:1979hx,Aragone:1981yn}. It says that the canonical way of putting fields on a curved background by replacing partial derivatives with covariant ones does not work for massless higher-spin fields. Indeed, in checking the gauge invariance of the action we have to commute derivatives, which brings the Riemann tensor:
      \begin{align}
      S&=\int \nabla\phi\nabla\phi+...\,, & \delta\phi&=\nabla\xi\,, & \delta S&=\int (\phi_{...})(\nabla_{.}\xi R_{\bullet\bullet,\bullet\bullet}+\xi \nabla_{.} R_{\bullet\bullet,\bullet\bullet})\,.
      \end{align}
Unlike low-spin examples, we find the full four-index Riemann tensor --- the structure that cannot be compensated by any modifications of the action/gauge transformations. For $s=1$ the action is manifestly gauge invariant, while for $s=3/2$ we find not the full Riemann tensor but its trace, the Ricci tensor, which allows to overcome the problem by going to supergravities.

The argument above makes use of the specific field variables and of the manifestly Lorentz covariant methods. Obviously, this is avoided by the light-cone approach. We will emphasize in Section \ref{subsec:covariant} that there exists in fact a two-derivative gravitational coupling of massless higher-spin fields to gravity  \cite{Bengtsson:1983pg,Bengtsson:1983pd,Bengtsson:1986kh}, which is not captured by covariant studies \cite{Boulanger:2006gr,Zinoviev:2008ck,Boulanger:2008tg}.

\paragraph{BCFW.} A relatively new no-go type result came from the BCFW approach \cite{Fotopoulos:2010ay,Benincasa:2007xk,Benincasa:2011pg,McGady:2013sga,Bengtsson:2016alt,Ponomarev:2016jqk}. However, higher-spin theories are clearly different from Yang-Mills theory and even gravity and are not expected to have an $S$-matrix that is analytic. Moreover, BCFW approach is essentially based on the assumption of certain behavior of amplitudes for infinite BCFW shifts. It is not a priori clear whether these assumptions can be justified in the higher-spin case. Some works towards weakening these assumptions include \cite{Fotopoulos:2010ay,Benincasa:2007xk,Benincasa:2011pg,McGady:2013sga,Bengtsson:2016alt}.
\paragraph{Three dimensions.} Massless higher-spin fields do not have local degrees of freedom in three-dimensions \cite{Blencowe:1988gj,Campoleoni:2010zq,Henneaux:2010xg,Gaberdiel:2012uj} and therefore the no-go theorems discussed above do not apply, see   \cite{Afshar:2013vka,Gonzalez:2013oaa} and references thereon for more detail.
\paragraph{AdS.} Another option to avoid the no-go theorems is to simply abandon the flat space and go to anti-de Sitter background \cite{Fradkin:1986qy,Vasiliev:1990en,Vasiliev:2003ev} since the no-go theorems discussed above were formulated for QFT's in flat space. 

\section{Living on Light-Front}
\label{sec:lightfront}
In this Section we review the light-cone approach to relativistic dynamics. Next, we discuss the classification of cubic vertices that results from the light-cone dynamics and confront it with the covariant methods. The main lesson is that there are more vertices in the light-cone approach. In particular there are two-derivative interaction vertices $s-s-2$ of a spin-$s$ field and a graviton, which can be called gravitational. The reader not interested in the somewhat boring details\footnote{Nice, pedagogical exposition of the light-cone approach can also be found in \cite{Metsaev:2005ar,Bengtsson:2012jm}.} can jump directly to Section \ref{subsec:covariant}. It is worth stressing that the Yang-Mills theory, when rewritten in the light-cone approach, is a theory of scalar fields in the adjoint of the global symmetry group. Similarly, gravity is a theory of two scalar fields with no symmetries like diffeomorphisms whatsoever.

\subsection{Basics}
Quantum field theory in flat space in its most rigorous definition requires a Hilbert space endowed with the unitary action of the Poincare algebra, i.e. the generators of Lorentz transformations $J^{AB}$ and translations $P^A$ should be realized as to obey:\footnote{It is convenient to choose the mostly plus convention for $\eta_{AB}$ and $A,B,...=0,...,d-1$.}
\begin{align}
[P^A,P^B]&=0\,,\\
[J^{AB},P^C]&=P^A\eta^{BC}-P^B\eta^{AC}\,,\\
[J^{AB},J^{CD}]&=J^{AD}\eta^{BC}-J^{BD}\eta^{AC}-J^{AC}\eta^{BD}+J^{BC}\eta^{AD}\,.
\end{align}
In free theory the generators are quadratic in the quantum fields and have to receive certain corrections when interactions are switched on.

Canonical quantization begins with postulating the canonical commutation relations of fields and momenta at some fixed time, which encodes the choice of the Cauchy surface for evolution. As was pointed out by Dirac \cite{Dirac:1949cp} there are different quantization schemes depending on the choice of the quantization surface.\footnote{Let us note that there are several different things that bear almost the same name: light-cone gauge, light-front (or light-cone) quantization and one can also combine the two by quantizing a theory on a light-front with the light-cone gauge imposed.} The difference is in the stability group that preserves the surface. The generators associated with the stability group, called {\it kinematical}, do not receive any quantum corrections and stay quadratic in the fields on the Cauchy surface. The left-over generators, called {\it dynamical}, do deform.

For the canonical equal time choice $t=t_0$ the stability subgroup of the Poincare group $ISO(3,1)$ ($ISO(d-1,1)$ in $d$ dimensions) consists of spacial rotations and translations, while boosts and time translations $P_0=H$ do not preserve the surface. Therefore, there are four generators ($d$ in the case of $d$ dimensions) that receive corrections due to interactions. 

The light-front is the light-like quantization surface. The canonical choice is $x^+=0$, so that $x^+$ is treated as the time direction and $H=P^-$ is the Hamiltonian.\footnote{In the light-front coordinates $A=+,-,a$, etc., $\eta^{+-}=\eta^{-+}=1$ and $\eta^{ab}=\delta^{ab}$. Also, in $4d$ one can replace $x^{1,2}$ with two complex conjugate variables $z$, $\bar{z}$, so that the metric is $2x^+x^-+2z\bar{z}$. }  As a result only $(d-1)$ generators need to be deformed, which is the least number possible. A somewhat unfortunate feature of any non-covariant quantization, including the light-cone one, is that due to the manifest Lorentz symmetry breaking we have to deal with many more generators whose total number is the same. The ten generators of $iso(3,1)$ can be split into kinematical (K) and dynamical (D) as follows:
\begin{align}
\text{kinematical}&: && P^{+}, P^{a}, J^{a+}, J^{+-}, J^{ab} &&: 7\\
\text{dynamical}&: && P^{-}, J^{a-} &&:3
\end{align}
The time evolution of any operator $G$ is determined by the Hamiltonian $\dot{G}=i[H,G]$. Therefore, if the Poincare algebra relations are satisfied at the initial light-cone time $x^+=0$, then they will be satisfied at all times. This has a useful consequence that some of the generators having explicit $x^+$ dependence
\begin{align} 
J^{-+}&= x^-\pl^+ - x^+ P^-\,,  & J^{a+}&=x^a\pl^+-x^+ \pl^a\,, 
\end{align}
should be declared to be kinematical, as we did above, since the dynamical part vanishes at $x^+=0$. The $x^+$-dependence can then be reconstructed by virtue of the equations of motion. 

Let us now list the commutation relations and the consequences thereof. A generator $G$ can be split $G=G_2+G_{\text{int}}$ into its free part $G_2$ and an interacting part $G_{\text{int}}$, the latter being absent for the kinematical generators. The kinematical generators are fixed once and for all times. As for dynamical generators the procedure is that there are commutators that simply constrain the dynamical generators to have certain dependence on the kinematical variables. Also, there are few other relations that represent nontrivial equations to be solved for the dynamical generators.

\paragraph{$\boldsymbol{[K,K]=K}$.} The kinematical generators do not receive any corrections, so this part stays unchanged and is of no use, which is why we list them here-below for completeness:
{\allowdisplaybreaks\besubeqs\begin{align}
[P^+,P^b]&=0\,, & [P^a,P^b]&=0\,,\\
[J^{-+},P^+]&=-P^+\,, &
[J^{-+},P^c]&=0\,, &
[J^{a+},P^+]&=0\,, \\
[J^{a+},P^c]&=-P^+ \delta^{ac}\,, &
[J^{ab},P^+]&=0\,, &
[J^{ab},P^c]&=P^a\delta^{bc}-P^b\delta^{ac}\,,\\
[J^{-+},J^{c+}]&=-J^{c+}\,, &
[J^{-+},J^{cd}]&=0\,, &
[J^{a+},J^{c+}]&=0\,, \\
[J^{a+},J^{cd}]&=\eta^{ac}J^{d+}-\eta^{ad}J^{c+}\,, &
[J^{ab},J^{cd}]&=\text{as usual}\,.
\end{align}\esubeqs}\noindent

\paragraph{$\boldsymbol{[K,D]=K}$.} This set of relations splits into two parts. First one is $[K,D]=0$-type relations that immediately restrict the dynamical generators. The second one are $[K,D]=K$-type commutators, which imply that the interacting part of $D$ commutes to the given $K$, i.e. $[K,D_{\text{int}}]=0$, which is due to $K_{\text{int}}=0$ and the right-hand side being taken into account by free fields, $[K_2,D_2]=K_2$.
\besubeqs\begin{align}
[P^+,P^-]&=0\,, &
[P^a,P^-]&=0\,, &
[J^{a+},P^-]&=P^a\,, \\
[J^{ab},P^-]&=0\,, &
[J^{a+},J^{c-}]&=\delta^{ac} J^{-+}-J^{ac}\,, &
[J^{a-}, P^+]&= P^a\,.
\end{align}\esubeqs

\paragraph{$\boldsymbol{[K,D]=D}$.} These relations are similar to the previous ones and constrain the dynamical generators to behave nicely under the light-front symmetries:
\besubeqs\begin{align}
[J^{-+}, P^-]&=P^-\,,&
[J^{-+},J^{c-}]&=J^{c-}\,, &
[J^{ab},J^{c-}]&=J^{a-} \delta^{bc}-J^{b-} \delta^{ac}\,,\\
[J^{a-},P^c]&=-P^- \delta^{ac}\,.
\end{align}\esubeqs

\paragraph{$\boldsymbol{[D,D]=0}$.} This class consists of the actual equations to be solved and constitutes the main problem of the light-cone approach:  
\begin{align}
[J^{a-},J^{c-}]&=0\,, &
[J^{a-},P^{-}]&=0\,.\label{hardequations}
\end{align}

\paragraph{Summary.} There are three dynamical generators: two boosts $J^{a-}$ and the Hamiltonian $H=P^-$:
\begin{align}
H=P^-&=H_2+ H\inter\,, & J^{a-}&=J^{a-}_2+x^a H\inter + J^{a-}\inter\,,
\end{align}
where we split them into the free and interacting parts and moreover symbolically extract the dependence of $J^{a-}$ on $H\inter$. The commutation relations imply that $H\inter$ is a centralizer of several kinematical generators:
\begin{align}\label{centerH}
[ H\inter,T]=0&: && T=P^+\,, P^a\,, J^{a+}\,, J^{ab} \,.
\end{align}
Initially, $ J^{a-}$ commutes only to $P^+$, $J^{a+}$. The shift by $x^a H\inter$ cancels $H$ in $[J^{a-},P^c]=-P^- \delta^{ac}$, which becomes the $[K,D]=0$-type relation for $J^{a-}\inter$. Therefore, we find
\begin{align}\label{centerJ}
[J^{a-}\inter,T]=0&: && T=P^+\,, P^a\,, J^{a+} \,.
\end{align}
The $[K,D]=D$-type relations, when written for the deformations, give
{\allowdisplaybreaks\begin{align}
[J^{-+},  H\inter]&=H\inter\,,&
[J^{-+}, J^{c-}\inter]&= J^{c-}\inter\,, \\
[J^{ab}, J^{c-}\inter]&= J^{a-}\inter \delta^{bc}- J^{b-}\inter \delta^{ac} \,.
\end{align}}\noindent
All the constraints above can be explicitly solved and one is left with \eqref{hardequations}, of which only $[H,J^{a-}]=0$ needs to be solved, as we explain below.

\subsubsection{Free Field Realization}
We have just discussed which commutation relations need to be solved. Further progress can only be made for specific theories. The general comment is that the quantization on the light-front leads to second-class constraints.\footnote{See very nice books \cite{Gitman:1990qh,Henneaux:1992ig} for quantization of field theories with constraints.} Indeed, the kinetic term $\tfrac12(\pl\phi)^2$, when written in the light-cone coordinates, $\pl^+\phi\pl^-\phi+\tfrac12 (\nabla\phi)^2$, is linear in the velocity $\pl^-\phi$ and hence the momenta, i.e. the primary constraints, cannot be solved for $\pl^-\phi$. Therefore, the bracket is the Dirac bracket.

From now on we confine ourselves to live in the four-dimensional world. The nice feature of the $4d$ world is that all massless spinning particles have two degrees of freedom, i.e. made of two scalar fields except for the spin-zero particle, which equals one scalar field. A spin-$s$ particle has two states with helicities $\pm s$ and can be described as two fields $\Phi^{\pm s}(x)$ that are complex conjugate. It is convenient to work with the fields that are Fourier transformed with respect to $x^-$ and transverse coordinates $x^a$:
\begin{align}
    \Phi(x,x^+)&=(2\pi)^{-\tfrac{d-1}2} \int e^{+i(x^-p^++x\cdot p)} \Phi(p,x^+)\, d^{d-1}p\,,\\
    \Phi(p,x^+)&=(2\pi)^{-\tfrac{d-1}2}  \int e^{-i(x^-p^++p\cdot x)} \Phi(x,x^+)\, d^{d-1}x\,.
\end{align}
In the $4d$ world the equal time commutation relations that follow from the Dirac bracket are:
\begin{align}\label{equaltime}
    [\Phi^{\mu}(p,x^+),\Phi^{\lambda}(q,x^+)]&=\delta^{\mu,-\lambda}\frac{\delta^{3}(p+q)}{2p^+}\,.
\end{align}
From now on we set $x^+=0$ and will omit the arguments in most of the cases. It is very easy to find the kinematical generators of the Poincare algebra in the Fourier space:\footnote{Following the light-cone commandments we rename $\beta=p^+$ otherwise the paper will not be understandable at all.}
\besubeqs\begin{align}
\hat{P}^+&=\beta\,, & \hat{P}&=p\,, & \hat{\bar{P}}&=\pb\,,\\
\hat{J}^{z+}&=-\beta\pfrac{\pb}\,, & \hat{J}^{\zb+}&=-\beta\pfrac{p}\,, & \hat{J}^{-+}&=-N_\beta-1=-\frac{\pl}{\pl \beta} \beta\,,\\
\hat{J}^{z\zb}&= N_p-N_\pb -\lambda\,,
\end{align}\esubeqs
where $N_p=p\pl_p$ is the Euler operator, {\it idem.} for $N_\pb$, $N_\beta$ and we sometimes use $\pl_\beta=\pl/\pl \beta$, etc. The generators are supposed to act on $\Phi^\lambda\equiv\Phi^\lambda_p\equiv\Phi^\lambda(\beta,p,\pb,x^+=0)$. The dynamical generators at the free level are:
\begin{align}
    H_2&=-\frac{p\pb}{\beta}\,, &&
    \begin{aligned}
        \hat{J}^{z-}_2&= \pfrac{\pb} \frac{ p\pb}{\beta} +p \pfrac{\beta} +\lambda\frac{p}{\beta}\,,\\
         \hat{J}^{\zb-}_2&= \pfrac{p} \frac{ p\pb}{\beta} +\pb \pfrac{\beta} -\lambda\frac{\pb}{\beta}  \,.     
    \end{aligned}
\end{align}
The Poincare charges can be built in a standard way:
\begin{align}
Q_\xi&= \int p^+\, d^{3}p\,\Phi^{-\mu}_{-p} O_\xi(p,\pl_p)\Phi^{\mu}_p\,,
\end{align}
where $O_\xi$ is the generator of the Poincare algebra associated with a Killing vector $\xi$. We draw reader's attention to the fact that the integration measure is $p^+$. The Poincare algebra is then realized via commutators
\begin{align}
    \delta_\xi \Phi^\mu(p,x^+)&= [\Phi^\mu(p,x^+),Q_\xi]\,.
\end{align}
Due to the nontrivial integration measure the conjugate operators are defined as
\begin{align}
    O^\dag &= -\frac{1}{p^+} O^T(-p) p^+\,,
\end{align}
where the transposed operator is defined via integration by parts as usual, e.g. $p^T=p$, $\pl_p^T=-\pl_p$. The generators of the Poincare algebra given above are Hermitian, $O^\dag=O$. In particular, we find $p^\dag=p$. With the help of \eqref{equaltime} and
\begin{align}\label{quadraticf}
     \delta_\xi \Phi^\mu(p,x^+)&=\frac12 O_\xi(p,\pl_p)  \Phi^\mu(p,x^+)+\frac12 O^\dag_\xi  \Phi^\mu(p,x^+)=O_\xi(p) \Phi^\mu_p
\end{align}
one can verify all the commutation relations: 
\begin{align}
    [Q_\xi,Q_\eta]&=Q_{[\xi,\eta]}\,, & [\delta_\xi,\delta_\eta]\Phi&=+\delta_{[\xi,\eta]}\Phi\,.
\end{align}
\eqref{quadraticf} follows from a more general formula for the action of $Q$ on an arbitrary functional $F[\Phi]$:
\begin{align}
    [F(\Phi),Q_\xi]&=\int dv\, O_\xi(v)\Phi^\nu_v\pfrac{\Phi^\nu_v} F[\Phi]\,,
\end{align}
which we will immediately apply to read off the constraints imposed by kinematical generators on the dynamical ones.

\subsubsection{Kinematical Constraints}
An appropriate ansatz for the Hamiltonian $H$ and dynamical boosts $J^{a-}$ reads:\footnote{The derivatives can also act both on $h$ and wave functions, which is equivalent to redefining $j^{a-}$. }
{\allowdisplaybreaks\besubeqs\begin{align}
    H&=H_2+\sum_n\int d^{3n}q\,\delta\left(\sum q_i\right) h_{\lambda_1...\lambda_n}^{q_1,...,q_n}\, \Phi^{\lambda_1}_{q_1}...\Phi^{\lambda_n}_{q_n}\,,\\
    J^{z-}&=J^{z-}_2+\sum_n\int d^{3n}q\, \deltas{\sum q_i}\left[ j_{\lambda_1...\lambda_n}^{q_1,...,q_n}-\frac{1}{n}h_{\lambda_1...\lambda_n}^{q_1,...,q_n}\left(\sum_j \pfrac{\bar{q}_j}\right)\right]\, \Phi^{\lambda_1}_{q_1}...\Phi^{\lambda_n}_{q_n}   \,,\\ 
    J^{\zb-}&=J^{\zb-}_2+\sum_n\int d^{3n}q\, \deltas{\sum q_i}\left[ \jb_{\lambda_1...\lambda_n}^{q_1,...,q_n}-\frac{1}{n} h_{\lambda_1...\lambda_n}^{q_1,...,q_n}\left(\sum_j \pfrac{q_j}\right)\right]\, \Phi^{\lambda_1}_{q_1}...\Phi^{\lambda_n}_{q_n}   \,,  
\end{align}\esubeqs}\noindent
where the delta function imposes the conservation of the total $q^+$ and transverse momenta $q,\qb$, which is a consequence of the translation invariance imposed by $P^a$ and $P^+$, \eqref{centerH}, \eqref{centerJ}. The rest of the kinematical generators imposes the following constraints:
{\allowdisplaybreaks\besubeqs\begin{align}\label{jap1}
    J^{a+}&: && \left(\sum_k \beta_k \pfrac{q^a_k}\right) h_{\lambda_1,...,\lambda_n}^{q_1,...,q_n}\sim0\,, \\
    J^{a+}&: && \left(\sum_k \beta_k \pfrac{q^a_k}\right) j_{\lambda_1,...,\lambda_n}^{q_1,...,q_n}\sim0\label{jap2}\,, && \text{same for } \jb\,,\\
    J^{z\zb}&: &&\left[\sum_k (N_{q_k}-N_{\qb_k})+\sum \lambda_k\right]h_{\lambda_1,...,\lambda_n}^{q_1,...,q_n}\sim0\,,\\
    J^{-+}&: && \sum_k \beta_k \pfrac{\beta_k} h_{\lambda_1,...,\lambda_n}^{q_1,...,q_n}\sim0\,,\\
    J^{-+}&: && \sum_k \beta_k \pfrac{\beta_k} j_{\lambda_1,...,\lambda_n}^{q_1,...,q_n}\sim0\,, && \text{same for } \jb\,,\\
    J^{z\zb}&: &&\left[\sum_k (N_{q_k}-N_{\qb_k})+\sum \lambda_k-1\right]j_{\lambda_1,...,\lambda_n}^{q_1,...,q_n}\sim0\,,\\
    J^{z\zb}&: &&\left[\sum_k (N_{q_k}-N_{\qb_k})+\sum \lambda_k+1\right]\jb_{\lambda_1,...,\lambda_n}^{q_1,...,q_n}\sim0 \,,   
\end{align}\esubeqs}\noindent
where $\sim0$ means an equality up to an overall delta-function $\delta^{d-1}(\sum q_k)$. 

In practice it is tedious to keep all delta-functions unresolved and it is more convenient to choose some independent momenta as basic variables. Moreover, \eqref{jap1}-\eqref{jap2} imply that everything depends on specific combinations of momenta $\mathbb{P}_{km}$:
\begin{align}
    J^{a+}&: && \sum_k \beta_k \pfrac{q^a_k}\sim0 && \Longrightarrow && \PP_{km}^a=q_k^a\beta_m-q^a_m\beta_k\,.
\end{align}
There are $N-2$ such independent variables for $N$-point function. In the $4d$ case we have
\begin{align}
   \PP_{km}&=q_k\beta_m-q_m\beta_k\,, & \PPb_{km}&=\qb_k\beta_m-\qb_m\beta_k \,.
\end{align}
Therefore, we assume that some $N-2$ variables out of all $\PP$'s have been chosen and
\begin{align}
    h_{\lambda_1...\lambda_n}(q_1,...,q_n)&=h_{\lambda_1...\lambda_n}(\PP_{km},\PPb_{km},\beta_k)\,,\\
    j_{\lambda_1...\lambda_n}(q_1,...,q_n)&=j_{\lambda_1...\lambda_n}(\PP_{km},\PPb_{km},\beta_k)\,, \qquad \text{same for } \jb\,.
\end{align}
The rest of the system of kinematical constraints can be rewritten as
{\allowdisplaybreaks\besubeqs\label{kinematics}\begin{align}
    J^{z\zb}&: &&\left[\PP \pfrac{\PP} -\PPb \pfrac{\PPb}+\sum \lambda_k\right]h_{\lambda_1,...,\lambda_n}^{q_1,...,q_n}\sim0\,,\\
    J^{-+}&: && \left[\PP \pfrac{\PP} +\PPb \pfrac{\PPb}+\sum_k \beta_k \pfrac{\beta_k}\right] h_{\lambda_1,...,\lambda_n}^{q_1,...,q_n}\sim0\,,\\
    J^{-+}&: && \left[\PP \pfrac{\PP} +\PPb \pfrac{\PPb}+\sum_k \beta_k \pfrac{\beta_k}\right]j_{\lambda_1,...,\lambda_n}^{q_1,...,q_n}\sim0\,,\\
    J^{-+}&: && \left[\PP \pfrac{\PP} +\PPb \pfrac{\PPb}+\sum_k \beta_k \pfrac{\beta_k}\right]\jb_{\lambda_1,...,\lambda_n}^{q_1,...,q_n}\sim0\,,\\
    J^{z\zb}&: &&\left[\PP \pfrac{\PP} -\PPb \pfrac{\PPb}+\sum \lambda_k-1\right]j_{\lambda_1,...,\lambda_n}^{q_1,...,q_n}\sim0\,,\\
    J^{z\zb}&: &&\left[\PP \pfrac{\PP} -\PPb \pfrac{\PPb}+\sum \lambda_k+1\right]\jb_{\lambda_1,...,\lambda_n}^{q_1,...,q_n}\sim0\,.
\end{align}\esubeqs}\noindent
The above conditions are very simple homogeneity constraints and need no further comments.

\subsubsection{Cubic Vertices}
The first nontrivial dynamical constraints arise at the cubic order. First of all, the kinematics of three $(d-1)$-dimensional momenta restricted by the conservation delta-function is very simple. There is one independent $\PP^a$ variable since $\PP^a_{12}=\PP^a_{23}=\PP^a_{31}$. Therefore, in $4d$ we have just $\PP$ and $\PPb$. It is advantageous to represent it in a manifestly cyclic-invariant way:
\begin{align}
    \PP^a_{12}&=...=\PP^a=\frac13\left[ (\beta_1-\beta_2)q^a_3+(\beta_2-\beta_3)q^a_1+(\beta_3-\beta_1)q^a_2\right]\,,\\
    \sigma_{123}\PP&=\PP\,, \qquad\qquad \sigma_{12}\PP=\sigma_{23}\PP=\sigma_{13}\PP=-\PP\,.
\end{align}
Therefore, $\PP$ belongs to the totally anti-symmetric representation of $S_3$. There is an identity that is of utter importance for the cubic approximation:
\begin{align}
    \sum_j \pfrac{q_j} \PP=0 \label{cubicmiracle}\,.
\end{align}
Also, at the three-point level we find
 \begin{align}
     \sum_i H_2(q_i)&=\frac{\PP \PPb}{\beta_1\beta_2\beta_3}= \frac{\PP \cdot \PP}{2\beta_1\beta_2\beta_3}\,.
\end{align}
Now we proceed to the dynamical constraints. The first one is $[H,J^{a-}]=0$ restricted to the cubic order in fields $\Phi$:
\begin{align}
    [H,J^{a-}]\Big|_3=[H_3,J^{a-}_2]-[J^{a-}_3,H_2]&=0\,,
\end{align}
which, after using the magic identity \eqref{cubicmiracle}, can be shown to lead to
\begin{align}
    \sum_i H_2(q_i) j_{3}&= \sum_i (\hat{J}^{z-}_2)^T\, h_3\,,&  \sum_i H_2(q_i) \jb_{3}&= \sum_i (\hat{J}^{\zb-}_2)^T\, \bar{h}_3\,,
\end{align}
where the transposed generators are
\begin{align}
    (\hat{J}^{z-}_2)^T&= -\frac{ q\qb}{\beta}\pfrac{\qb} -q \pfrac{\beta} +\lambda\frac{q}{\beta}\,, &
    (\hat{J}^{\zb-}_2)^T&= - \frac{ q\qb}{\beta}\pfrac{q} -\qb \pfrac{\beta} -\lambda\frac{\qb}{\beta}  \,.      
\end{align}
Now one can make an appropriate ansatz for $h_3$ that solves the kinematical constraints \eqref{kinematics}, act with $J_2^T$ and read off $j_3$ and $\jb_3$ up to possible redefinitions. The most general case is studied in Appendix \ref{app:cubics}, while below we simply quote the representation given by Metsaev in \cite{Metsaev:1991nb,Metsaev:1991mt}. The first results on cubic interactions of HS fields were obtained in \cite{Bengtsson:1983pg,Bengtsson:1983pd,Bengtsson:1986kh} in a slightly different base.

At the interaction level there is always a problem of fixing the field redefinitions. The light-cone approach is not free of this ambiguity too. At the cubic order redefinitions allow one to eliminate powers of $\PP \PPb\sim H_2$, but not each of the two separately. Therefore, the most natural choice of the redefinition frame is to have purely holomorphic vertices. It is worth stressing that this is not the most natural choice in the covariant approaches. The vertices are \cite{Metsaev:1991nb,Metsaev:1991mt}:
\besubeqs\label{famouscubic}\begin{empheq}[box=\fbox]{align}
    h_{\lambda_1,\lambda_2,\lambda_3}&= C^{\lambda_1,\lambda_2,\lambda_3} \frac{\PPb^{\lambda_1+\lambda_2+\lambda_3}}{\beta_1^{\lambda_1}\beta_2^{\lambda_2}\beta_3^{\lambda_3}}+\bar{C}^{-\lambda_1,-\lambda_2,-\lambda_3} \frac{\PP^{-\lambda_1-\lambda_2-\lambda_3}}{\beta_1^{-\lambda_1}\beta_2^{-\lambda_2}\beta_3^{-\lambda_3}}\,,\\
    j_{\lambda_1,\lambda_2,\lambda_3}&=+\frac23 C^{+\lambda_1,+\lambda_2,+\lambda_3} \frac{\PPb^{+\lambda_1+\lambda_2+\lambda_3-1}}{\beta_1^{+\lambda_1}\beta_2^{+\lambda_2}\beta_3^{+\lambda_3}}\Lambda^{\lambda_1,\lambda_2,\lambda_3}\,,\\
    \jb_{\lambda_1,\lambda_2,\lambda_3}&=-\frac23\bar{C}^{-\lambda_1,-\lambda_2,-\lambda_3} \frac{\PP^{-\lambda_1-\lambda_2-\lambda_3-1}}{\beta_1^{-\lambda_1}\beta_2^{-\lambda_2}\beta_3^{-\lambda_3}}\Lambda^{\lambda_1,\lambda_2,\lambda_3}\,,
\end{empheq}\esubeqs
where
\begin{align}
    \Lambda=\beta_1(\lambda_2-\lambda_3)+\beta_2(\lambda_3-\lambda_1)+\beta_3(\lambda_1-\lambda_2)\,.
\end{align}
Here $C^{\lambda_1,\lambda_2,\lambda_3}$ and $\bar{C}^{-\lambda_1,-\lambda_2,-\lambda_3}$ are two sets of coupling constants which are a priori independent. For dimensional reasons we have to introduce a parameter $l_P$ with the dimension of length as to compensate for the higher powers of momenta, as was noted as early as \cite{Bengtsson:1983pg,Bengtsson:1983pd,Bengtsson:1986kh}:
\begin{align}
    C^{\lambda_1,\lambda_2,\lambda_3}&= (l_P)^{\lambda_1+\lambda_2+\lambda_3-1} c^{\lambda_1,\lambda_2,\lambda_3}\,,&& \text{same for }\bar{C}\,.
\end{align}
In higher-spin theories the parameter will be naturally associated with the Planck length as the Einstein-Hilbert vertex is a part of the set above and corresponds to $C^{2,2,-2}$.

The light-cone locality implies that the powers of $\PP$, $\PPb$ must be non-negative or whenever $\sum \lambda_i=0$ we should have $\lambda_i=0$. The latter is due to the fact that $j^{a-}_3$  has one power of $\PP$ or $\PPb$ less. The exception is when all $\lambda_i=0$, which is the scalar self-interaction vertex, since it leads to $j_3=0$, which is implied in \eqref{famouscubic}.

Let us stress that the light-cone approach deals only with physical degrees of freedom, so the light-cone gauge is a unitary gauge, but it is not an on-shell method. Nevertheless, there is a striking relation between the on-shell amplitude methods and the light-cone approach \cite{Ananth:2012un,Akshay:2014qea,Bengtsson:2016jfk}. One can introduce 
\begin{equation}
\label{9a11}
|i] = \frac{2^{1/4}}{\sqrt{\beta_i}}\left(\begin{array}{c}
 \bar q_i \\
  -\beta_i
\end{array}\right)=2^{1/4}\left(\begin{array}{c}
 \bar q_i \beta_i^{-1/2} \\
 - \beta^{1/2}_i
\end{array}\right)\,,
\end{equation}
so that the basic building blocks of cubic vertices can be found in 
\begin{equation}
\label{9a1}
[i | j] = \sqrt{\frac{2}{\beta_i\beta_j}}\PPb_{ij}\,,
\end{equation}
and analogously one can define $|i\rangle$. As a result, the cubic vertices, i.e. Hamiltonian density $h_3$, can be rewritten in a more suggestive form:
\begin{align}\label{ampvertices}
    C^{s_1,s_2,s_3}[12]^{s_1+s_2-s_3}[23]^{s_2+s_3-s_1}[13]^{s_1+s_3-s_2}+c.c.\,,
\end{align}
which are the usual amplitudes for three helicity fields \cite{Benincasa:2007xk,Benincasa:2011pg}.

\subsection{Light-Cone vs. Covariant Vertices}
\label{subsec:covariant}
On one hand, the general formula for cubic vertices \eqref{famouscubic} is given above. On another hand, the classification of cubic vertices in covariant approaches is also available.\footnote{There is a vast literature on cubic vertices in covariant approaches. We give a minimalistic list of references \cite{Fradkin:1986qy,Buchbinder:2006eq,Zinoviev:2008ck,Boulanger:2008tg,Manvelyan:2010je,Bekaert:2010hp,Zinoviev:2010cr,Fotopoulos:2010ay,Taronna:2011kt,Joung:2013nma} that allows one to trace all the initial results and further developments by following references therein/thereon, the accent being put on the diversity of approaches. For our purposes it is sufficient to confront the classification of the light-cone vertices \cite{Metsaev:1993ap,Metsaev:2005ar,Metsaev:2007rn} with some of the covariant results \cite{Boulanger:2008tg,Bekaert:2010hp}. } Remarkably, by confronting the light-cone vertices and covariant ones we observe a mismatch in the number of local interactions, see also \cite{Benincasa:2011pg,Benincasa:2007xk, Conde:2016izb,Sleight:2016xqq}, which is due to the difference between locality in light-cone and covariant approaches.  

In the light-cone approach, the vertices can be arranged by the number of derivatives for a given triplet of spins $s_1-s_2-s_3$. The power counting is easy in the light-cone approach too: one counts the total power of the transverse momenta, or of $\PP^a$, which is the same. Therefore, vertices \eqref{famouscubic} have $|\lambda_1+\lambda_2+\lambda_3|$ derivatives, where we note that the helicities can be negative.
In covariant approaches one can distinguish between the following classes of vertices, though this classification is incomplete:

\begin{description}
    \item[Current Interaction.] The $C^{0,0,s}$ vertex has $s$ derivatives and corresponds to the usual current interaction where a spin-$s$ current $\phi_0\pl^s\phi_0$ built of two scalar fields is contracted with the Fronsdal field $\phi_s$. This is the simplest vertex that involves one higher-spin field and for $s=1$ corresponds to the current interaction while for $s=2$ to the coupling $T_{\mu\nu}g^{\mu\nu}$ of the stress-tensor to gravity.
    
    \item[Non-abelian Vertices.] For every spin the vertex $C^{s,s,-s}$ has $s$ derivatives and drives the non-abelian deformation of the gauge algebra in the covariant approach \cite{Bekaert:2010hp}. In $d>4$ there can be more than one non-abelian self-interaction, but in $4d$ this seems to be the only one. In particular, $C^{1,1,-1}$ is the Yang-Mills vertex and $C^{2,2,-2}$ is the Einstein-Hilbert vertex. Having such vertices activated is important for non-triviality of the theory. There is a covariant vertex $s-s-2$ with  $(2s-2)$ derivatives that can be called gravitational, but it certainly cannot result from $\pl\rightarrow \nabla$ replacement in the action due to its higher derivative nature for $s>2$. 
    
    \item[Abelian Vertices.] It is also possible to construct the $(s_1+s_2+s_3)$-derivative vertex $C^{s_1,s_2,s_3}$. It does not induce any deformation of the gauge algebra and therefore cannot be used as a seed of any interesting theory, while such vertices can be required for consistency at the quartic order. This indeed happens for higher-spin theory, but does not happen for Yang-Mills and Einstein theory, where $F^3$ and $R^3$ vertices can be dropped (or have an independent coupling constant in front of them). 
    
\end{description}

As a result, there is a mismatch between the covariant and the light-cone dictionaries. Indeed, $s_1-s_2-s_3$ vertex in $4d$ can have $s_1+s_2+s_3-2\min(s_1,s_2,s_3)$ or $s_1+s_2+s_3$ derivatives \cite{Metsaev:2005ar}, i.e. one can have two vertices at most. On contrary, the light-cone vertices exist for any triplet of helicities, i.e. there can be up to four independent complex vertices \eqref{famouscubic}. When the reality condition is imposed, $C=\bar{C}$, this number still reduces to three vertices at most. For example,  there exists an exceptional series of vertices $C^{s',s,-s}$, $s>s'$, that have less derivatives (transverse momenta) and is absent in covariant approaches.  In particular, this exceptional series contains a two-derivative $C^{2,s,-s}$ gravitational vertex!\footnote{The existence of such an vertex was stressed in \cite{Bengtsson:2014qza}, though it is certainly present in \cite{Bengtsson:1986kh,Metsaev:1991nb,Metsaev:1991mt,Metsaev:1993ap}.} The $s=2$ case corresponds to the usual Einstein-Hilbert vertex and does not look strange anymore.

The existence of such vertex seems paradoxical in view of the simplest no-go result, known as the Aragone-Deser argument \cite{Aragone:1979hx}. As we discussed in Section \ref{sec:notsonogo} the argument is explicitly Lorentz covariant and is formulated in terms of specific field content, Fronsdal fields, rather than in terms of physical degrees of freedom and therefore is avoided by the light-cone approach.

More generally, within the covariant approaches the statement that some interaction does not exist depends heavily on the field content. Few examples include: local electromagnetic interactions mediated by $A_\mu$ will look non-locally in terms of $F_{\mu\nu}$; the formulation of self-dual fields may require an infinite number of auxiliary fields \cite{Berkovits:1997wj}; there may be the need for some compensator or other auxiliary fields, see e.g. \cite{Pasti:2009xc}; a seeming breaking of Lorentz symmetries might be needed, e.g. \cite{Henneaux:1988gg}.

Another result that seems to be in tension with the existence of the two-derivative $s-s-2$ vertex is the Weinberg low energy theorem. As we discussed in Section \ref{sec:notsonogo}, the light-cone approach seems to avoid the assumptions of the theorem.  

It is worth stressing that the existence of the strange low-derivative vertex is not a unique feature of the light-cone approach and is also seen via amplitude techniques \cite{Benincasa:2011pg,Benincasa:2007xk}, as \eqref{ampvertices} reveals.

\subsection{Quartic Analysis and Beyond}
The main result of the cubic approximation is the list of all possible cubic vertices that can be used for constructing any theory. As is usual for the cubic approximation, the coefficients $C^{s_1,s_2,s_3}$ and $\bar{C}^{s_1,s_2,s_3}$ in front of the cubic vertices are completely free and will be fixed by the quartic analysis, which we will now proceed to. 

After an appropriate ansatz for $j_n$ and $h_n$ that solves the kinematical constraints \eqref{kinematics} is chosen, one has to solve
\begin{align}
    [J^{a-}_2,H_n]-[H_2,J^{a-}_n]&=\sum_{\substack{i,j>2\\i+j=n}}[H_i^{\vphantom{a}},J^{a-}_j]\,.
\end{align}
The right-hand side contains the source that is made of the structures that are supposed to have been already found at orders lower than $n$. On evaluating the commutators on the left-hand side we discover an operator that commutes to the delta-functions that impose momenta conservation:
\begin{align}
\tilde{J}^{a-}_2(q_i)=\tilde{J}^{a-}(q_i,\pl_{q_i})= J^{a-}_2(q_i)^T-H_2(q_i)\frac{1}{n}\left(\sum_j \pfrac{q^a_j}\right)
\,.
\end{align}
For further convenience let us denote
\begin{align}
\label{defHJ}
\JJJ^{a-}_2&=\sum_i \tilde{J}^{a-}_2(q_i)\,, &
\HHH_2&=\sum_i H_2(q_i)\,.
\end{align}
Now the main equation can be rewritten as
\begin{align}\label{maineq}
  \HHH_2\,  j^{a-}_n&=\JJJ^{a-}_2[ h_n]+\sum_{\substack{i,j>2\\i+j=n}}[H_i^{\vphantom{a}},J^{a-}_j]\,.
\end{align}
It is evident that the equation can formally be solved for $j_n$ by dividing both sides by $\HHH_2$. However, when we go on-shell $\HHH_2$ is the sum of $p^-$ components and will vanish. Therefore, the crucial requirement is to adjust the right-hand side of \eqref{maineq} as to make it be proportional to $\HHH_2$. Then, we can safely divide by $\HHH_2$ and get some local $j_n$. If such a solution is found we can all go to the beach because it can be shown, see Appendix \ref{app:jj}, that $[J^{a-},J^{b-}]=0$ contains no new relations and holds true automatically.  

\paragraph{Parity transformations.} The parity is a useful symmetry as well as its breaking is. The parity transformations are defined as
\begin{align}
    P(\Phi^\lambda)&=\Phi^{-\lambda}\,, & P(C^{\lambda_i})&=\bar{C}^{-\lambda_i}\,, & P(\PP)&=\PPb\,.
\end{align}
If we are interested in the unitary higher-spin theory we have to impose $C=\bar{C}$, otherwise the Hamiltonian is complex. For the chiral theories this condition will be violated.

\section{Complete Chiral Higher-Spin Theory}
\label{sec:selfdual}
In this section we will show that there exists a complete chiral higher-spin theory in flat space. Also we will elaborate on the solution obtained by Metsaev in \cite{Metsaev:1991nb,Metsaev:1991mt}, with the technical details placed in Appendix \ref{app:insym}.

The starting point is the ansatz for $H_4$ and $J^{a-}_4$ that solves the kinematical constraints \eqref{kinematics} and is free of delta-functions:
\begin{align}
H_4&= \int h_4(\PP_{12},\PP_{34};\beta_i)\Phi^{\lambda_1}_{q_1}...\Phi^{\lambda_4}_{q_4}\,,\\
J_4&=\int j_4(\PP_{12},\PP_{34};\beta_i)\Phi^{\lambda_1}_{q_1}...\Phi^{\lambda_4}_{q_4}-\frac14 h_4(\PP_{12},\PP_{34};\beta_i) \left(\sum \pfrac{\qb_j}\right)\Phi^{\lambda_1}_{q_1}...\Phi^{\lambda_4}_{q_4}\,,\\
\bar{J}_4&=\int \jb_4(\PP_{12},\PP_{34};\beta_i)\Phi^{\lambda_1}_{q_1}...\Phi^{\lambda_4}_{q_4}-\frac14 h_4(\PP_{12},\PP_{34};\beta_i) \left(\sum \pfrac{q_j}\right)\Phi^{\lambda_1}_{q_1}...\Phi^{\lambda_4}_{q_4}\,.
\end{align}
The consistency condition to be solved at the quartic order is
\begin{align}\label{maineq4}
  \HHH_2\,  j^{a-}_4&=\JJJ^{a-}_2[ h_4]+[H_3^{\vphantom{a}},J^{a-}_3]\,.
\end{align}
For definiteness, let us consider the component of this equation with $a=z$, i.e. the one for $j$ while the equation for $\jb$ is similar. Following Metsaev \cite{Metsaev:1991nb,Metsaev:1991mt}, we note that
both $\HHH_2\,  j^{a-}_4$ and $\JJJ^{a-}_2[ h_4]$, if non-zero, are at least
linear in $q$. On the other hand, the contribution from the anti-holomorphic part of
$H_3$ (which we denote $H_3(\bar{\mathbb{P}})$) to $[H_3^{\vphantom{a}},J_3]$ is $q$-independent and thus has to vanish on its own:
\begin{align}
\label{8sep1}
    [H_3^{\vphantom{a}}(\bar{\mathbb{P}}),J_3] &=0\,,
    && [H_3^{\vphantom{a}}(\PP),\Jb_3] =0\,.
\end{align}
{\it Therefore, the parts of the quartic consistency condition \eqref{maineq4} that have $CC$ and $\bar{C}\bar{C}$ structure constants form a system of equation that is decoupled from $h_4$ and $j_4$!} We called these parts holomorphic. The $C\bar{C}$-part of $[H_3,J^{a-}_3]$ does couple to $h_4$ and $j^{a-}_4$ and its analysis is a real challenge:
\begin{align}\label{hardeq}
    \HHH_2\,  j_4&=\JJJ_2[ h_4]+[H_3^{\vphantom{a}}(\PP),J_3]\,, &
    \HHH_2\,  \jb_4&=\bar{\JJJ}_2[ h_4]+[H_3^{\vphantom{a}}(\PPb),\Jb_3]\,.
\end{align}
The holomorphic equations \eqref{8sep1} impose a strong constraint on coupling constants $C^{\lambda_1,\lambda_2,\lambda_3}$ and in fact can be used to fix all of them in terms of a single coupling constant provided some reasonable conditions on the theory are imposed. Certainly one can fulfill the holomorphic constrains by selecting the abelian vertices only, which gives an infinitely many of not so interesting solutions. The complete classification of solutions is still lacking. Also, there is a series of solutions where a single spin-$s$ field couples to graviton and Yang-Mills field, see also Section \ref{subsec:universal}. Generally, one can find solutions if $C^{\lambda_1,\lambda_2,\lambda_3}$ is sufficiently sparse and does not include non-abelian self-interactions of higher-spin fields, $C^{s,s,-s}$, as they force one to introduce all higher-spin fields together. This is coherent with the studies \cite{Fradkin:1986ka,Maldacena:2011jn,Alba:2013yda,Boulanger:2013zza,Stanev:2013qra,Alba:2015upa} of uniqueness of higher-spin symmetries in the context of AdS/CFT.

The solution found by Metsaev  \cite{Metsaev:1991nb,Metsaev:1991mt} has a remarkably simple form:
\begin{align}
\label{Ruscoup}
C^{\lambda_1,\lambda_2,\lambda_3} =\frac{(l_P)^{\lambda_1+\lambda_2+\lambda_3-1}}{\Gamma(\lambda_1+\lambda_2+\lambda_3)}\,.
\end{align}
where $l_P$ has to have a dimension of length and, since there is graviton in the multiplet, it is natural to identify $l_P$ with the Planck length. Even though the final result (\ref{Ruscoup}) is very simple, its derivation is not obvious. To make our discussion more self-contained we present in Appendix \ref{app:insym} a relatively simple and explicit derivation of Metsaev's formula also highlighting some of its important features.

The remarkable property of the $4d$ light-cone approach is that the consistency equations split into two decoupled systems of $CC$ and $\bar{C}\bar{C}$ equations \eqref{8sep1} for the structure constants of the cubic action and an additional system \eqref{hardeq} that contains $h_4$ and $j^{a-}_4$. The latter is the system to be solved for $h_4$ and $j^{a-}_4$ while the source involves $C\bar{C}$. A crucial observation is that one can simply set $\bar{C}=0$ and hence $h_4=0$, $j^{a-}_4=0$ is a solution of \eqref{hardeq}. Obviously, setting $h_{n>3}=0$ and $j^{a-}_{n>3}=0$ together with $C$ from \eqref{Ruscoup} provides a complete solution to all orders! The only feature is that the Hamiltonian is complex and for that reason the theory in non-unitary.

For completeness we write below the full Hamiltonian of the chiral higher-spin theory and an action that can be obtained by the Legendre transform:
\besubeqs\label{sdhs}\begin{align}
H&=\int \Phi^{-\lambda}_{-q} \frac{q\qb}{\beta}\Phi^\lambda_q + \int \frac{(l_p)^{\lambda_1+\lambda_2+\lambda_3-1}}{\Gamma(\lambda_1+\lambda_2+\lambda_3)} \frac{\PPb^{\lambda_1+\lambda_2+\lambda_3}}{\beta_1^{\lambda_1}\beta_2^{\lambda_2}\beta_3^{\lambda_3}}\Phi^{\lambda_1}_{q_1}\Phi^{\lambda_2}_{q_2}\Phi^{\lambda_3}_{q_3}\delta^{3}(q_1+q_2+q_3)\,,\\
S&=-\int \pl_A\Phi^{-\lambda} \pl^A\Phi^\lambda + \int \frac{(l_p)^{\lambda_1+\lambda_2+\lambda_3-1}}{\Gamma(\lambda_1+\lambda_2+\lambda_3)} \frac{\PPb^{\lambda_1+\lambda_2+\lambda_3}}{(\pl^+_1)^{\lambda_1}(\pl^+_2)^{\lambda_2}(\pl^+_3)^{\lambda_3}}\Phi^{\lambda_1}\Phi^{\lambda_2}\Phi^{\lambda_3}\,,
\end{align}\esubeqs
where the sum over all helicities $\lambda$ is assumed and the fields in the last line $\Phi^\lambda(x)$ carry full space-time dependence. The derivatives are to act on the corresponding fields and in the last line \be\PPb=\frac13\left[ (\pl^+_1-\pl^+_2)\plb_3+(\pl^+_2-\pl^+_3)\plb_1+(\pl^+_3-\pl^+_1)\plb_2\right]\,. \ee

\paragraph{Tree-Level four-point Amplitude.} In this paragraph we show that the four-point scattering amplitude
in the chiral theory given by (\ref{sdhs}) vanishes. The total  $s$-channel
exchange between external fields with 
helicities $\lambda_1$, $\lambda_2$, $\lambda_3$ and $\lambda_4$
is
\begin{equation}
 \label{amp1}
 {\cal A}_s =\sum_{\omega}\frac{1}{(\lambda_1+\lambda_2+\omega-1)!} \frac{\bar{\mathbb{P}}^{\lambda_1+\lambda_2+\omega}_{12}}{\beta_1^{\lambda_1}\beta_2^{\lambda_2}} \frac{1}{(q_1+q_2)^2} 
 \frac{1}{(\lambda_3+\lambda_4-\omega-1)!} \frac{\bar{\mathbb{P}}^{\lambda_3+\lambda_4-\omega}_{34}}{\beta_3^{\lambda_3}\beta_4^{\lambda_4}}.
\end{equation}
Performing summation and adding contributions from other channels we find
\begin{align}
\notag
 {\cal A} =\frac{1}{2(\Lambda -2)!\prod_{i=1}^4\beta_i^{\lambda_i}}
\Big( &\frac{\bar{\mathbb{P}}_{12}\bar{\mathbb{P}}_{34}}{(q_1+q_2)^2}[(\bar{\mathbb{P}}_{34}-\bar{\mathbb{P}}_{12})^{\Lambda -2}-(\bar{\mathbb{P}}_{34}+\bar{\mathbb{P}}_{12})^{\Lambda -2}]\\
\notag
 + &\frac{\bar{\mathbb{P}}_{13}\bar{\mathbb{P}}_{24}}{(q_1+q_3)^2}[(\bar{\mathbb{P}}_{24}-\bar{\mathbb{P}}_{13})^{\Lambda -2}-(\bar{\mathbb{P}}_{24}+\bar{\mathbb{P}}_{13})^{\Lambda -2}]\\
  \label{amp2}
 +
 & \frac{\bar{\mathbb{P}}_{14}\bar{\mathbb{P}}_{23}}{(q_1+q_4)^2}[(\bar{\mathbb{P}}_{14}-\bar{\mathbb{P}}_{23})^{\Lambda -2}-(\bar{\mathbb{P}}_{14}+\bar{\mathbb{P}}_{23})^{\Lambda -2}]\Big),
\end{align}
where $\Lambda = \sum_{i=1}^4 \lambda_i$.
For on-shell momenta one has
\begin{align}
 (q_i+q_j)^2 &= -\frac{2}{\beta_i\beta_j}\mathbb{P}_{ij}\bar{\mathbb{P}}_{ij}\,, && \sum_{j=1}^4\frac{\mathbb{P}_{ij}\bar{\mathbb{P}}_{jk}}{\beta_j}=0.
\end{align}
These identities allow one to show that 
\begin{equation}
E\equiv \frac{\bar{\mathbb{P}}_{12}\bar{\mathbb{P}}_{34}}{(q_1+q_2)^2}
 =-\frac{\bar{\mathbb{P}}_{13}\bar{\mathbb{P}}_{24}}{(q_1+q_3)^2} =
 \frac{\bar{\mathbb{P}}_{14}\bar{\mathbb{P}}_{23}}{(q_1+q_4)^2}\,.
\end{equation}
Also, using 
\begin{align}
\notag
2A\equiv &\;  \bar{\mathbb{P}}_{12}+\bar{\mathbb{P}}_{34}= -\bar{\mathbb{P}}_{14}+\bar{\mathbb{P}}_{23}\,,\\
 \label{3may4}
  2 B\equiv & \;\bar{\mathbb{P}}_{13}-\bar{\mathbb{P}}_{24}= \bar{\mathbb{P}}_{34}-\bar{\mathbb{P}}_{12}\,,\\
  \notag
 2C\equiv & \;  \bar{\mathbb{P}}_{14}+\bar{\mathbb{P}}_{23}= -\bar{\mathbb{P}}_{13}-\bar{\mathbb{P}}_{24}\,
 \end{align}
 we find  
 \begin{align*}
  {\cal A}=\frac{E}{2(\Lambda -2)!\prod_{i=1}^4\beta_i^{\lambda_i}}
  \big( (2B)^{\Lambda-2}-(2A)^{\Lambda-2} -(2B)^{\Lambda-2}+(2C)^{\Lambda-2}+(2A)^{\Lambda-2}-(2C)^{\Lambda-2} \big)=0.
 \end{align*}
Therefore, the four-point amplitude is zero and we expect all higher-point tree-level amplitudes to vanish as well.

\section{Quartic Hamiltonian of Unitary Higher-spin Theory}
\label{sec:quarticH}

In the previous section  we considered the consistency condition
(\ref{maineq4}), which we repeat here for convenience 
\begin{align}\label{12sep1}
  \HHH_2\,  j^{a-}_4&=\JJJ^{a-}_2[ h_4]+[H_3^{\vphantom{a}},J^{a-}_3],
\end{align}
for $z=a$. Following Metsaev, we showed that they contain an
independent sector, which involves only the couplings $C^{\lambda_1,\lambda_2,\lambda_3}$ of the
chiral part of the Hamiltonian and allow to fix them up to
inessential normalisation factors, see (\ref{9sep3}) in Appendix \ref{app:insym}. 
The remaining consistency conditions can be solved trivially
by setting to zero the coupling constants $\bar C^{-\lambda_1,-\lambda_2,-\lambda_3}$
of the conjugated 
Hamiltonian. This choice of the cubic Hamiltonian is consistent
on its own and thus defines a complete higher-spin theory \eqref{sdhs}.

Analogously, one can start from (\ref{12sep1}) with $a=\bar z$
and fix the coupling constants $\bar C^{-\lambda_1,-\lambda_2,-\lambda_3}$ of the anti-chiral Hamiltonian. Setting to zero the remaining coupling constants
one obtains a consistent anti-chiral higher-spin theory.

As is expected the Hamiltonian of the chiral theory is not Hermitian. Bearing in mind the fact that there should be a one-parameter family of higher-spin theories in $AdS_4$ \cite{Giombi:2011kc} with the self-dual limits as extremal cases, we expect that there should exist a one-parameter family of higher-spin theories in flat space too. The starting point is to take $Ce^{i\gamma}$ and $\bar{C}e^{-i\gamma}$ as new couplings with $C$ and $\bar{C}$ given by the Metsaev formula. The chiral theories arise in the $e^{\pm i\gamma}\rightarrow\infty$ limits. In particular, there should exist a unitary higher-spin theory, which we are after. In the unitary theory we should have $C^{\lambda_1,\lambda_2,\lambda_3}=\bar C^{-\lambda_1,-\lambda_2,-\lambda_3}$, i.e. the theory is parity invariant.
Then, 
\begin{equation*}
    [H_3^{\vphantom{a}}({\mathbb{P}}),J^{z-}_3]\sim C \bar{C}\mathbb{P}\bar{\mathbb{P}} \ne 0
\end{equation*}
generates a non-vanishing contribution to (\ref{12sep1})
with $a=z$. This implies that the theory cannot be truncated
at the level of cubic vertices and requires higher order 
interaction terms. This story is very much parallel to the gravity case where the deformation procedure does not stop at the cubic level.

We recall that $\HHH_2$ is just an algebraic operator
that acts by multiplying by $\sum_i H_2(q_i)$, see (\ref{defHJ}). 
This means that (\ref{12sep1}) can always be solved formally for
$j_4^{a-}$ no mater what are the contributions from other terms.
This, however, requires to divide by $\sum_i H_2(q_i)$, which 
vanishes when all particles go on-shell. 
The only way to avoid this singularity is to require that 
the right hand side of (\ref{12sep1}) is proportional to 
$\sum_i H_2(q_i)$. In other words, we have to solve the 
equation
\begin{equation}
\label{12sep2}
\Big(\JJJ^{a-}_2[ h_4]+[H_3^{\vphantom{a}},J^{a-}_3]\Big)\Big|_{\sum_i H_2(q_i)=0}=0
\end{equation}
for $h_4$. Once the solution is found, we can simply solve
(\ref{12sep1}) for $j_4$ without producing a singularity.

In the remaining part of this section we solve (\ref{12sep2})
for $h_4$ in the case of spin zero self-interaction. To do that
we make a  general local ansatz, expressed in terms of independent
variables and fix free coefficients by requiring that (\ref{12sep2})
is satisfied. As independent variables parametrising dependence of
$h^4$ on transverse momenta we take $\mathbb{P}_{12}$,
$\mathbb{P}_{34}$, $\bar{\mathbb{P}}_{12}$ and $\bar{\mathbb{P}}_{34}$. Explicitly in the scalar case one has
{\allowdisplaybreaks\begin{align}
 [H_3,J_3^{z-}]=\sum_{\omega =0}\frac{3(-)^\omega\omega}{4} &(C^{00\omega})^2\Big[ \;\frac{\beta_1-\beta_2}{\beta_1+\beta_2}
 \frac{\bar{\mathbb{P}}_{12}^{\omega-1}\mathbb{P}_{34}^\omega}{(\beta_1+\beta_2)^\omega
 (\beta_3+\beta_4)^\omega}+
 \frac{\beta_3-\beta_4}{\beta_3+\beta_4}
 \frac{\bar{\mathbb{P}}_{34}^{\omega-1}\mathbb{P}_{12}^\omega}{(\beta_1+\beta_2)^\omega
 (\beta_3+\beta_4)^\omega}
 \notag
 \\
+ &\;\frac{\beta_1-\beta_3}{\beta_1+\beta_3}
 \frac{\bar{\mathbb{P}}_{13}^{\omega-1}\mathbb{P}_{24}^\omega}{(\beta_1+\beta_3)^\omega
 (\beta_2+\beta_4)^\omega}+
 \frac{\beta_2-\beta_4}{\beta_2+\beta_4}
 \frac{\bar{\mathbb{P}}_{24}^{\omega-1}\mathbb{P}_{13}^\omega}{(\beta_1+\beta_3)^\omega
 (\beta_2+\beta_4)^\omega}\\
 \label{27may1}
 +
 &\;\frac{\beta_1-\beta_4}{\beta_1+\beta_4}
 \frac{\bar{\mathbb{P}}_{14}^{\omega-1}\mathbb{P}_{23}^\omega}{(\beta_1+\beta_4)^\omega
 (\beta_2+\beta_3)^\omega}+
 \frac{\beta_2-\beta_3}{\beta_2+\beta_3}
 \frac{\bar{\mathbb{P}}_{23}^{\omega-1}\mathbb{P}_{14}^\omega}{(\beta_1+\beta_4)^\omega
 (\beta_2+\beta_3)^\omega}\Big]\,.\notag
 \end{align}}\noindent
Taking into account that $\JJJ^{z-}_2$ raises the homogeneity
degree in $\bar q$ by one and keeps the homogeneity degree in 
$q$ unchanged, we conclude that, if at all possible, one should
be able to solve (\ref{12sep2}) 
for each $\omega$ separately and use the ansatz of the form $h_4^{[\omega]} \sim \mathbb{P}^{\omega-1}\bar{\mathbb{P}}^{\omega-1}$.
In other words, $\omega$ provides a grading associated with the
number of transverse derivatives.
Moreover, from  lower-$\omega$ cases we learned that one
can find $h_4$'s compensating each line of (\ref{27may1}) 
separately. 
Using these observation we found a solution for a general $\omega$
to be
\begin{align}
   h_4&=\frac32\sum_\omega (-)^{\omega+1}(C^{0,0,\omega})^2h_4^{[\omega]}+ (\{1234\}\to \{1324\}) + (\{1234\}\to \{1423\})\,,
\end{align}
where
\begin{align*}
 h_4^{\omega}=&\frac{-1}{4}\frac{\beta_1\beta_2\beta_3\beta_4}{(\beta_1+\beta_2)^2(\beta_3+\beta_4)^2}s\sum_{n=1}^{\omega-1}
 \frac{\omega!}{n!(\omega-n)!}\Big(\frac{\bar{\mathbb{P}}_{12}\mathbb{P}_{34}}{(\beta_1+\beta_2)(\beta_3+\beta_4)}\Big)^{n-1}
 \Big(\frac{{\mathbb{P}}_{12}\bar{\mathbb{P}}_{34}}{(\beta_1+\beta_2)(\beta_3+\beta_4)}\Big)^{\omega-n-1}\\
 &\qquad+\sum_{n=0}^{\omega-1}\frac{\omega!}{n!(\omega-n)!}
 \Big(\frac{\bar{\mathbb{P}}_{12}\mathbb{P}_{34}+{\mathbb{P}}_{12}\bar{\mathbb{P}}_{34}}{(\beta_1+\beta_2)(\beta_3+\beta_4)}\Big)^{n}
 \Big(-\frac{1}{4}\frac{\beta_1-\beta_2}{\beta_1+\beta_2}\frac{\beta_3-\beta_4}{\beta_3+\beta_4} \Big)^{\omega-n} s^{\omega-n-1}\,,
\end{align*}
and following \cite{Metsaev:1991nb} we introduced an analog of the $s$ Mandelstam variable:  
\begin{equation*}
    s=\frac{\mathbb{P}_{12}{\bar{\mathbb{P}}}_{12}}{\beta_1\beta_2}+
    \frac{\mathbb{P}_{34}{\bar{\mathbb{P}}}_{34}}{\beta_3\beta_4}\,.
\end{equation*}
This gives the $0-0-0-0$ part of the quartic Hamiltonian of the unitary higher-spin theory. Note that the solution does not have transverse momenta in denominators and should be regarded as perturbatively local.

\section{Higher-Spin Equivalence Principle}
\label{sec:HSequivalence}
The Weinberg low-energy theorem \cite{Weinberg:1964ew} implies the conservation of the electric charge and the equivalence principle. However, it is too restrictive for massless higher-spin fields to have long-range interactions, see also \cite{Bekaert:2010hw}. As we already noted in Section \ref{sec:notsonogo} the theorem does not formally apply to the light-cone vertices. 

In this Section, after discussing few lower spin examples, we will prove that Yang-Mills and gravitational interactions of higher-spin fields exhibit certain universality. In particular, the equivalence principle extends to higher-spin fields as well. The Metsaev solution discussed above obeys the equivalence principle too. 

\subsection{Examples of Low Spin Fields}
What we try to see below is the conditions that arise at the quartic level when some set of cubic vertices is activated, i.e. to probe the holomorphic constraints \eqref{8sep1} that decouple from $H_4$ and $J_4$, but, as we have seen, can restrict couplings.

\paragraph{Scalar Cubed Theory.} This is the simplest and somewhat trivial example:
\begin{align}
h_3&=\Phi^0\Phi^0\Phi^0\left[C^{0,0,0}+\text{c.c.}\right]\,, & J_3&=0\,.
\end{align}
Thanks to $J_3=0$ the commutator $[J_3,H_3]$ vanishes identically revealing that the cubic vertex provides a self-consistent theory and solves \eqref{maineq4}, which is expected, of course.

\paragraph{Yang-Mills theory.} For the case of spin-one self-interaction we have to have a colored set of fields since $\PP$ is totally anti-symmetric. Therefore, we introduce some anti-symmetric structure constants $f^{abc}$ and let fields carry additional indices too, $\Phi^\lambda_a$. The cubic vertex reads
\begin{align}
h_3&=f^{abc}\Phi^1_a\Phi^1_b\Phi^{-1}_c\left[\frac{\PPb_{12} C^{1,1,-1}\beta_3}{\beta_1\beta_2}+\text{c.c.}\right]\,.
\end{align}
After summing over cyclic permutations we find that \eqref{8sep1} is satisfied provided the Jacobi identity for the structure constants is true.

\paragraph{Yang-Mills theory coupled to Scalar Matter.} It is also interesting to see how the Yang-Mills fields can couple to matter.\footnote{This example was also discussed in \cite{Metsaev:1991nb,Metsaev:1991mt}, as well as the scalar-tensor theory below.} To this effect we add a one-derivative $0-0-1$ vertex, where the current built of the scalar fields couples to the Yang-Mills field:
\begin{align}
h_3&=\left[\frac{\beta_3 }{\beta_1\beta_2}f^{abc}\Phi^1_a\Phi^1_b\Phi^{-1}_c{\PPb_{12} C^{1,1,-1}}+
T^{aij}\Phi^1_a\Phi^0_i\Phi^{0}_j C^{0,0,1}\frac{\PPb_{23}}{\beta_1}
+\text{c.c.}\right]\,.
\end{align}
Interestingly, after symmetrizing over the permutations \eqref{8sep1} implies $C^{1,1,-1}=C^{0,0,1}$, i.e. the coupling constants must be equal.

\paragraph{Pure Gravity.} In the case of pure gravity we inject the Einstein-Hilbert two-derivative cubic vertex, i.e. $C^{2,2,-2}\neq0$, while all other constants are zero:
\begin{align}
h_3=\Phi^2\Phi^2\Phi^{-2}\left[\frac{\PPb_{12}^2\beta_3^2 C^{2,2,-2}}{\beta_1^2\beta^2_2}+\text{c.c.}\right]
\end{align}
Then the holomorphic part \eqref{8sep1} of the commutator $[J_3,H_3]$ can be found to identically vanish after symmetrizing over permutations of all four legs, which, at this order, just tells us that gravity might be a consistent theory.

\paragraph{Higher-derivative Gravity.} From the covariant approach it is known that one can add a six-derivative  $R^3$-type vertex, the resulting theory being consistent. In the light-cone approach we start with
\begin{align}
h_3=\Phi^2\Phi^2\Phi^{-2}\left[\frac{\PPb_{12}^2\beta_3^2 C^{2,2,-2}}{\beta_1^2\beta^2_2}\right]+
\Phi^2\Phi^2\Phi^{2}\left[\frac{\PPb_{12}^6 C^{2,2,2}}{\beta_1^2\beta^2_2\beta_3^2}\right]
+\text{c.c.}\,.
\end{align}
In the commutator one finds two types of $CC$ terms:
\begin{align}
\eqref{8sep1} \sim (...)C^{2,2,-2}C^{2,2,-2}+(...)C^{2,2,-2}C^{2,2,2}\,,
\end{align}
which vanish independently after symmetrizing over the four legs. Therefore, the $R^3$ vertex can be added with an arbitrary coefficient, which is to be expected from the covariant approaches.

\paragraph{Gravity plus Scalar Matter.} A different situation is with the scalar-tensor theory, which in addition to gravity contains a two-derivative vertex that couples the scalar field stress-tensor to gravity:
\begin{align}
h_3=\Phi^2\Phi^2\Phi^{-2}\left[\frac{\PPb_{12}^2\beta_3^2 C^{2,2,-2}}{\beta_1^2\beta^2_2}\right]+
\Phi^0\Phi^0\Phi^{2}\left[\frac{\PPb_{12}^2 C^{0,0,2}}{\beta_3^2}\right]
+\text{c.c.}
\end{align}
In this case the vanishing of \eqref{8sep1} imposes a single constraint:
\begin{align}
C^{2,2,-2}=C^{0,0,2}\,,
\end{align}
i.e. the scalar field coupling equals to that of the gravity ---  the equivalence principle.

\paragraph{Einstein-Yang-Mills Theory.} We can also try to couple a spin-one field to gravity, i.e. to activate the $C^{2,1,-1}$ vertex:
\begin{align}
h_3=\Phi^2\Phi^2\Phi^{-2}\left[\frac{\PPb_{12}^2\beta_3^2 C^{2,2,-2}}{\beta_1^2\beta^2_2}\right]+
\Phi^1\Phi^{-1}\Phi^{2}\left[\frac{\PPb_{12}^2 C^{1,-1,2}\beta_2}{\beta_1\beta_3^2}\right]
+\text{c.c.}
\end{align}
As before the vanishing of \eqref{8sep1} imposes a single constraint:
\begin{align}
C^{2,2,-2}=C^{1,-1,2}\,,
\end{align}
i.e. the equivalence principle for a Maxwell field.

\subsection{Universality of Gravity and Yang-Mills}
\label{subsec:universal}
Even before attempting to look for a complete theory we can ask a simpler question: what happens if we have a higher-spin field which is coupled to gravity or the Yang-Mills theory.

Generalizing the low-spin examples above, we can take a spin-$s$ field and a spin-one Yang-Mills field and turn on $C^{s,-s,1}$ in addition to the Yang-Mills interaction itself. Then, vanishing of the holomorphic terms in $[H_3,J_3]$ implies that all higher-spin fields couple universally to spin-one:
\begin{align*}
s-s-1&:&    C^{s,-s,1}&=C^{1,1,-1}=g\,.
\end{align*}
The same exercise for the gravitation interaction, i.e. with $C^{2,2,-2}$ and $C^{2,s,-s}$ switched on implies that all higher-spin fields couple universally to spin-two:\footnote{Technically, what one does is to take $h_3$ with $C^{2,2,-2}$, $C^{2,s,-s}$ vertices and then to symmetrize over the fields in \eqref{8sep1}. The outcome is proportional to a complicated kinematical factor and $C^{2,s,-s}(C^{2,2,-2}-C^{2,s,-s})$, which leads to the result and similarly in the case of the Yang-Mills interaction.}
\begin{align*}
s-s-2&:&    C^{s,-s,2}&=C^{2,2,-2}=g\,l_{p}\,.
\end{align*}
The fact that the strength of the backreaction from higher-spin fields on gravity must be the same for all spins $s=0,1,2,3,4,...$ is a reincarnation of the equivalence principle which, as it turns out, holds true for fields of any spin.\footnote{We do not consider fermionic higher-spin fields in this paper, but undoubtedly they have to follow the same law.} 

The higher-spin equivalence principle also implies that there is a system made of graviton and a spin-$s$ field with only the Einstein-Hilbert $C^{2,2,-2}$ and gravitational $C^{s,-s,2}$ vertices switched on that solves the holomorphic constraints \eqref{8sep1}. Therefore, this solution explicitly avoids the Aragone-Deser argument in the light-cone approach and suggests that it may be possible to put higher-spin fields on more general backgrounds. However, \eqref{8sep1} is a necessary condition and an obstruction can come from the rest of the constraints \eqref{hardeq} and higher orders.

It should be noted that the Weinberg low-energy theorem, if applied literally to the higher-spin case, does imply that all couplings should be equal but it simultaneously imposes a too restrictive conservation law that can only be obeyed by the scattering processes that simply permute the particles' momenta. Pessimistically, this should then be seen later in the light-cone approach too. Optimistically, the Weinberg theorem can be avoided by the light-cone approach.

\section{Conclusions and Discussion}
\label{sec:conclusions}
We pointed out that due to the holomorphic splitting of the Poincare algebra consistency relations there exists a complete chiral higher-spin theory in $4d$ flat space. Such a theory provides a counterexample to a widespread belief that higher-spin interactions are impossible in the Minkowski space. However, the theory is non-unitary.

While the chiral theory is an encouraging result, we expect the unitary higher-spin theory to exist too. Its derivation requires more efforts since the Poincare deformation procedure does not stop at the cubic order. We have fixed a part of the quartic Hamiltonian that determines an infinite series of the quartic contact vertices of the scalar field. This can be thought of as the Minkowski space counterpart of the AdS result obtained recently in \cite{Bekaert:2014cea,Bekaert:2015tva}. In particular the flat space quartic action shares some features with its $AdS_4$ cousin: it is naively non-local in having an unbounded order in derivatives arranged into a series of positive powers of the transverse momenta. However, there are no wild non-localities of type $1/\square$ or $1/p_i\cdot p_j$, which would trivialize the deformation procedure \cite{Barnich:1993vg}. Such non-localities arise in some of the covariant studies \cite{Fotopoulos:2010ay}, but not in the others \cite{Taronna:2011kt}. Formally, the quartic scalar self-interaction drops off the Noether procedure at this order since scalar field does not feature its own gauge parameter. The equation for the quartic scalar vertex is a part of the quintic Noether consistency conditions.  

The mild non-localities we observed are to be expected since higher-spin theories are not power-counting renormalizable and coupling conspiracy is the only way for them to be quantum consistent, i.e. to have an infinitely-many of Slavnov-Taylor identities as a result of a clever fine-tuning of higher-derivative interactions. The light-cone locality requirement is not to have transverse momenta in denominators, otherwise any deformation can be formally extended to higher-orders \cite{Barnich:1993vg}. The quartic $0-0-0-0$ coupling we found is local in this sense.

One of the surprises of the light-cone approach as compared to covariant methods is the existence of an additional, exceptional, series of cubic vertices which contains the two-derivative gravitational interactions of higher-spin fields. These vertices are also seen by the amplitude methods.

As was observed by Metsaev, at the quartic order the Poincare algebra consistency relations split into the three parts, two of which do not involve the quartic generators at all but do impose restrictions on possible couplings. Some mild assumptions on the spectrum of a theory are needed, otherwise there are infinitely many solutions, some of which might be of interest too in the context of conformal higher-spin fields, \cite{Metsaev:unpublished}, or gravitational interactions. We gave a simple derivation of the solution found by Metsaev in Appendix \ref{app:insym}. 

The latter was impossible to see by covariant methods for the two reasons: (i) the holomorphic decomposition of the Poincare algebra consistency relations is essentially not Lorentz covariant; (ii) the solution we are interested in requires the exceptional vertices to be present. Therefore, at least for some of the problems the covariant methods turn out to be too restrictive (or at least an appropriate set of auxiliary or compensator fields is requires and still unknown). Let us note that this solution rules out indirectly consistent higher-spin theories in flat space that are manifestly Lorentz covariant. Here it is worth stressing that any QFT resulting from the light-cone approach is Poincare invariant by definition.

These results are the $4d$ Minkowski counterparts of the higher-spin algebra uniqueness theorems \cite{Fradkin:1986ka,Maldacena:2011jn,Boulanger:2013zza,Alba:2013yda,Stanev:2013qra,Alba:2015upa} proved recently in the context of AdS/CFT, which imply the same statement for massless higher-spin fields in $AdS$ or higher-rank conserved tensors in the dual CFT picture. The latter results are even more restrictive because the higher-spin algebras can be shown to be generated by free conformal fields.

A more general idea that we would like to pursue is to establish a relation between higher-spin theories in flat and AdS spaces. In particular, it would be interesting to see if there exists a flat limit in some sense (naive limit of vanishing cosmological constant should not work beyond the cubic level \cite{Boulanger:2008tg}). It is tempting to propose that the light-cone approach is a suitable framework for such a limit to be smooth. Indeed, one finds many similarities between higher spins in flat and AdS spaces to support this idea. 

Firstly, it is the vanishing of the scalar self-coupling, which is a sensible interaction. It comes as a surprise in flat space. We would like to stress that this is consistent with the $AdS$-lift of this theory where the absence of $\phi^3$ is required by the $AdS/CFT$ correspondence \cite{Klebanov:2002ja,Sezgin:2003pt,Giombi:2009wh}. In the critical vector model $\langle \sigma\sigma\sigma\rangle|_{d=3}=0$ and therefore the bulk coupling is expected to be zero at $d=4$. Meanwhile, in the free vector model $\langle \varphi^2\varphi^2\varphi^2\rangle\neq0$, but the bulk vertex is extremal, \cite{Sezgin:2003pt,Giombi:2009wh}, therefore the bulk coupling should approach zero near four-dimensions, and indeed it does \cite{Bekaert:2014cea}.

Secondly, in flat space the cubic action is given by the simple Metsaev solution $\Gamma(\lambda_1+\lambda_2+\lambda_3)^{-1}$. Later \cite{Bekaert:2014cea} the same pattern was observed for $s-0-0$ vertices in $AdS_4$ higher-spin theory and conjectured \cite{Skvortsov:2015pea} to be the same for all $s_1-s_2-s_3$ with the explicit proof given in \cite{Sleight:2016dba} in the course of reconstructing the complete cubic action of the Type-A higher-spin theory. A new piece of evidence may come from the twistorial approach to conformal higher-spin theory \cite{Haehnel:2016mlb}, if the three-point functions of the unitary truncation turn out to be the same.

Thirdly, both in AdS and flat cases we see the option of having a consistent parity-violating theory, whose extreme limit is the chiral theory presented in the paper. This is just an observation in flat space, while the AdS counterpart is well supported by the existence of Chern-Simons matter theories \cite{Giombi:2011kc}. It would be interesting to construct the chiral higher-spin theory in AdS too, which is supposed to terminate at the cubic level contrary to the unitary higher-spin theories. In this regard, it is worth mentioning that there is a one-parameter family of boundary conditions \cite{Giombi:2013yva}:
    \begin{align}
        e^{+i\gamma} C&=e^{-i\gamma }\bar{C}\,, && e^{i\gamma}=\sqrt{\frac{k+iN}{k-iN}}\,,
    \end{align}
where $k$ is the Chern-Simons level and $N$ is the number of matter fields. The two standard limits are $C=\pm \bar{C}$ and correspond to ordinary and alternate boundary conditions. It is interesting that there are two extremal cases where $e^{\pm i\gamma}$ goes to infinity and therefore imposes $C=0$ or $\bar{C}=0$. Clearly, in the bulk such a limit results in a self-dual higher-spin theory, while its interpretation from the CFT side is unclear \cite{Soojong}. The simplicity of the self-dual $AdS_4$ higher-spin theory that we expect is based on its flat space cousin. 

Another fruitful direction to go is to extend the $4d$ quartic results to higher-dimensions. In particular, it is interesting to see if there exists a phenomenon similar to the holomorphic factorization of the Poincare algebra at the quartic order that allows to fix the cubic action before encountering any problems at the quartic level. Lower dimensions $d=5,...$ should be of more interest due to the specific structures on the Wigner little group. In particular in $d=5$ the relevant algebra is $su(2)$ and therefore the spinning degrees of freedom should be governed by $hs(\lambda)$ \cite{Vasiliev:1989re,Feigin} which is familiar from the $3d$ higher-spin studies \cite{Campoleoni:2010zq,Henneaux:2010xg,Gaberdiel:2012uj}. The case of $AdS_5/CFT_4$ higher-spin duality can be richer owing to the existence of doubletons \cite{Gunaydin:1984fk,Gunaydin:1984wc}, i.e. massless conformal fields of arbitrary spin. For any $j=0,\tfrac12,1,\tfrac32,...$ there should exist a higher-spin dual, called Type-A,B,C,... that computes the correlators of the primaries that are bilinear in spin-$j$ doubletons. This should work classically, while there can be some obstructions at the quantum level for $j>1$ \cite{Gunaydin:2016amv}, see also \cite{Giombi:2013fka,Giombi:2014iua,Giombi:2014yra} for the details of the approach. The relevant one-parameter family of higher-spin algebras was discussed in \cite{Fradkin:1989md,Fernando:2009fq,Boulanger:2011se,Govil:2013uta,Manvelyan:2013oua,Joung:2014qya}, while the relation to $hs(\lambda)$ is manifest in the quasiconformal approach of \cite{Fernando:2009fq,Govil:2013uta}.

Another direction along the lines of recent studies \cite{Caron-Huot:2016icg} is to try to solve the quartic consistency relations for a stringy spectrum of fields, i.e. instead of massless higher-spin fields one can try to add massive higher-spins fields and to see what are the options for the multiplets that are consistent with the Poincare algebra. It is not difficult to see that the presence of an at least one massive higher-spin field will require the spin in the multiplet to be unbounded from above as in the massless case. The detailed classification of such multiplets is absent. 

Lastly, we attempted to construct the unitary higher-spin theory in flat space. Even though we found only the quartic scalar self-interaction and not the full quartic Hamiltonian we at least have not faced any obstructions. Moreover, in the $4d$ light-cone approach higher-spin fields are not that different from the scalar one. In this respect our result gives us hope that reconstruction of the full quartic Hamiltonian is also possible. On another hand, the no-go results, especially those that were obtained within the BCFW approach that is closer to the light-cone one as compared to covariant methods, still suggest that the light-cone analysis can face certain difficulties as well. It would be interesting to establish this in future.

\section*{Acknowledgement}
We are indebted to our light-cone coach Ruslan Metsaev without whose patient explanations this work would have been impossible. We would like to thank Lars Brink and Anders Bengtsson for useful discussions and correspondence. Discussions with Maxim Grigoriev, Philipp H\"{a}hnel, Igor Klebanov, Kirill Krasnov, Tristan McLoughlin, Soo-Jong Rey, Ivo Sachs, Arkady Tseytlin and Yuri Zinoviev are very much appreciated. We also would like to thank Nicolas Boulanger, Ruslan Metsaev and Arkady Tseytlin for the very useful comments on the draft. The work of E.S. was supported by the Russian Science Foundation grant 14-42-00047 in association with Lebedev Physical Institute and by the DFG Transregional Collaborative Research Centre TRR 33 and the DFG cluster of excellence ”Origin and Structure of the Universe”.
The work of D.P. was supported by the ERC Advanced grant No.290456. E.S. and D.P. also acknowledge a kind hospitality at the program ``Higher Spin Theory and Duality" MIAPP, Munich (May 2-27, 2016) organized by the Munich Institute for Astro- and Particle Physics (MIAPP).
Some of the results of this paper were reported at 'Higher-spin Theory and Duality' conference held at MIAPP, Munich, 23-25, May 2016. D.P. also would like to thank Arnold Sommerfeld Center for Theoretical Physics for the hospitality.

\begin{appendix}
\renewcommand{\thesection}{\Alph{section}}
\renewcommand{\theequation}{\Alph{section}.\arabic{equation}}
\setcounter{equation}{0}\setcounter{section}{0}

\section{Most General Form of Cubic Vertices}
\label{app:cubics}
\setcounter{equation}{0}
We write the most general ansatz that solves all the kinematical constraints, i.e. have the correct homogeneity in $\PP$, $\PPb$ and $\beta$'s:
\begin{align}
    h_{\lambda_1,\lambda_2,\lambda_3}&= C^{\lambda_1,\lambda_2,\lambda_3} \frac{\PPb^{\lambda_1+\lambda_2+\lambda_3}}{\beta_1^{\lambda_1}\beta_2^{\lambda_2}\beta_3^{\lambda_3}}F\left[\frac{\PP\PPb}{\beta_1\beta_2},\frac{\beta_1}{\beta_2}\right]\,,
\end{align}
where $F[x,y]$ is a priori an arbitrary function of two arguments. Also we solved explicitly for the momenta conservation, so that $\beta_3$ is unnecessary.
Applying $J_2$ we find:
\begin{align}
    \frac{1}{h}\sum J_2^T h&=\frac{1}{F}\frac{\PP\left (O[F]-2\frac{\Lambda}{\beta_2} F(x,y)\right)}{3 \beta_2^2 y (y+1)}\,,\\
    \frac{1}{h}\sum \bar{J}_2^T h&=\frac{1}{F}\frac{\PPb}{3 \beta_2^2 y (y+1)} O[F]\,,\\
    O[F]&=-3 y (y+1) F^{(0,1)}(x,y)-x (y-1) F^{(1,0)}(x,y)\,,
\end{align}
where in the last live we defined the differential operator $O$ that contributes both to $J_2 h_3$ and $\Jb_2 h_3$. It has a zero mode that is responsible for the field redefinitions: $f(\tfrac{x^3 y}{(y+1)^2})$. Therefore, it is convenient to rewrite the ansatz as
\begin{align}
h_{\lambda_1,\lambda_2,\lambda_3}&= C^{\lambda_1,\lambda_2,\lambda_3} \frac{\PPb^{\lambda_1+\lambda_2+\lambda_3}}{\beta_1^{\lambda_1}\beta_2^{\lambda_2}\beta_3^{\lambda_3}}F\left[\frac{\PP\PPb}{(\beta_1\beta_2\beta_3)^{2/3}},\frac{\beta_1}{\beta_2}\right]\,.
\end{align}
We should not worry about fractional powers. Whenever needed they can always be compensated by the $y$-dependence. Now the operators simplify a bit (simple derivative remains)
\begin{align}
    \frac{1}{h}\sum J_2^T h&=\frac{1}{F}\frac{\PP\left (-3 y (y+1) F^{(0,1)}(x,y)-2\frac{\Lambda}{\beta_2} F(x,y)\right)}{3 \beta_2^2 y (y+1)}\,,\\
    \frac{1}{h}\sum \bar{J}_2^T h&=\frac{1}{F}\frac{-\PPb F^{(0,1)}(x,y)}{\beta_2^2}\,.
\end{align}
This provides the most general solution for the generators at the cubic level:
\begin{align}
    j_3&=\left[\frac{\PP \PPb}{\beta_1\beta_2\beta_3}\right]^{-1} \sum J_2^T h_3\,, &
    \jb_3&=\left[\frac{\PP \PPb}{\beta_1\beta_2\beta_3}\right]^{-1}\sum \bar{J}_2^T h_3\,.
\end{align}
The redefinitions correspond to adding a multiple of $H_2 \sim \PP\PPb$. Therefore, the solutions can be made purely (anti)-holomorphic in $\PP$ and $\PPb$, which is the choice made by Metsaev and quoted in the main text.

\section{Triviality of $\boldsymbol{[J,J]=0}$}
\label{app:jj}
\setcounter{equation}{0}
The last and the most difficult part of the commutation relations $[J^{a-},J^{b-}]=0$ is always true\footnote{We are indebted to Ruslan Metsaev for claiming that this fact should be true.} provided $J^{a-}$ is solved for in terms of $H$ at a given order. Indeed, at order $n$ and with all kinematical constraints already solved for we have two equations:
\begin{align}
    [J_2^{a-},H_n]+\sum_{i+j=n; i,j>1}[J_i^{a-},H_j]&=[H_2,J^{a-}_n]\,,\\
    [J_2^{a-},J^{b-}_n]-(ab)+\sum_{i+j=n; i,j>1}[J_i^{a-},J_j^{-b}]&=0\,.
\end{align}
We note that the action of $H_2$ in the first equation is always algebraic and is therefore invertible off-shell, i.e. outside the zero-energy surface $\sum E_i=0$. Let us start at the cubic order where we have 
\begin{align}
    [J_2^{a-},H_3]&=[H_2,J^{a-}_3]\,, &
    [J_2^{a-},J^{b-}_3]-(ab)&=0\,.
\end{align}
In order to check that the last equality is automatically true we use the invertibility of $[H_2,\bullet]$:
\begin{align}
    [J_2^{a-},J^{b-}_3]-(ab)&=0\,, && \Longleftrightarrow && 
    [H_2,[J_2^{a-},J^{b-}_3]]-(ab)=0  \,.  
\end{align}
Using the Jacobi identity and $[H_2,J^{a-}_2]=0$ we have the following chain of implications
\begin{align}
&[H_2,[J_2^{a-},J^{b-}_3]]-(ab)=[J_2^{a-},[H_2,J^{b-}_3]]-(ab)=\\
&=[J_2^{a-},[J_2^{b-},H_3]]-(ab)=
-[H_3,[J^{a-}_2,J^{b-}_2]]=0\,,
\end{align}
where in the last step we used the algebra relations at one order less. At the quartic order we find
\begin{align}
    [H_2,[J_2^a,J_4^b]]-(ab)+[H_2,[J_3^a,J_3^b]]=\\
    [J_2^a,[H_2,J_4^b]-[J_3^a,[H_2,J_3^b]]-(ab)=\\
    [J_2^a,[J_2^b,H_4]]+[J_2^a,[J_3^b,H_3]]-[J_3^a,[H_2,J_3^b]]-(ab)=\\
    0-[H_3,[J_2^a,J_3^b]]-[J_3^a,[J_2^b,H_3]]-(ab)=[H_3,[J_3^a,J_2^b]]-(ab)=0
\end{align}
where we used several times the cubic order relations $[J_2,J_3]=0$ and $[H_2,J_3]=[J_2,H_3]$. The general proof follows the same logic, but is a bit boring.

\section{Metsaev Solution}
\label{app:insym}
\setcounter{equation}{0}
First step is to evaluate explicitly the commutator in (\ref{8sep1}), which results in:
\begin{align}
\sum_{\omega}\rm{Sym}\Big[\frac{(\lambda_1+\omega-\lambda_2)\beta_1-(\lambda_2+\omega-\lambda_1)\beta_2}{\beta_1+\beta_2}C^{\lambda_1,\lambda_2,\omega}C^{\lambda_3,\lambda_4,-\omega}\bar{\mathbb{P}}_{12}^{\lambda_1+\lambda_2+\omega-1}\bar{\mathbb{P}}_{34}^{\lambda_3+\lambda_4-\omega}\Big]=0\,,
  \label{8sep2}
\end{align}
where $\rm{Sym}$
is a complete symmetrisation, which originates from contraction with
 $\Phi_{q_i}^{\lambda_i}$. This symmetrisation is essential:
if it had been omitted, the solution (\ref{Ruscoup}) would have been
lost. The 
expression appearing in brackets in (\ref{8sep2}) is manifestly symmetric with respect to 
permutations $1\leftrightarrow 2$ and $3\leftrightarrow 4$. To achieve complete 
symmetry
one has to add  five other non-trivial permutations
\begin{align*}
   6\cdot  \{1,2,3,4\} \to \{1,2,3,4\}+\{1,3,2,4\} + \{1,4,2,3\} +\{3,4,1,2\} +
    \{2,4,1,3\} + \{2,3,1,4\}\,.
\end{align*}

Provided momentum conservation is taken into account,
there are five independent variables
among 
$\bar{\mathbb{P}}_{ij}$ and $\beta_i$. So, generically, to
solve an equation of the form (\ref{8sep2}) one would need to express the left hand-side
in terms of five independent variables and
then solve it for all values of these variables.
It, however,  turns out, that the left hand side of (\ref{8sep2})
can be expressed in terms of $\bar{\mathbb{P}}_{ij}$ only, among which 
only three are independent. This can be seen if we group the 
term in brackets in (\ref{8sep2}) and the one  obtained
by the permutation $\{1,2,3,4\} \to \{3,4,1,2\}$ and relabeling $\omega\to- \omega$. Summing them, we find that the  $\beta$-dependence cancels
\begin{align}
 \notag
 \frac{(\lambda_1+\omega-\lambda_2)\beta_1-(\lambda_2+\omega-\lambda_1)\beta_2}{\beta_1+\beta_2}\bar{\mathbb{P}}_{34}+ \frac{(\lambda_3-\omega-\lambda_4)\beta_3-(\lambda_4-\omega-\lambda_3)\beta_4}{\beta_3+\beta_4}\bar{\mathbb{P}}_{12}\\
 \label{intp}
 =(\lambda_1-\lambda_2)\bar{\mathbb{P}}_{34}+(\lambda_3-\lambda_4)\bar{\mathbb{P}}_{12}+
 \frac{\omega}{2}(\bar{\mathbb{P}}_{13}-\bar{\mathbb{P}}_{23}+\bar{\mathbb{P}}_{24}-\bar{\mathbb{P}}_{14})\,.
 \end{align}
In terms of  independent variables (\ref{3may4}), equation 
(\ref{8sep2}) can be rewritten in a more suggestive form
{\allowdisplaybreaks
  \begin{align}
 \notag
 \sum_{\omega} \Big[& \big((\lambda_1-\lambda_2+\lambda_3-\lambda_4)A+(\lambda_1-\lambda_2-\lambda_3+\lambda_4)B-2\omega C\big)\\
 \notag
 &\qquad \qquad \qquad  \qquad  \qquad  C^{\lambda_1,\lambda_2,\omega}C^{\lambda_3,\lambda_4,-\omega}
 (A-B)^{\lambda_1+\lambda_2+\omega-1}(A+B)^{\lambda_3+\lambda_4-\omega-1}\\
 \notag
 +&\big((-\lambda_1+\lambda_3+\lambda_2-\lambda_4)B+(-\lambda_1+\lambda_3-\lambda_2+\lambda_4)C+2\omega A\big)\\
 \notag
&\qquad \qquad \qquad  \qquad  \qquad  C^{\lambda_1,\lambda_3,\omega}C^{\lambda_2,\lambda_4,-\omega}
 (B-C)^{\lambda_1+\lambda_3+\omega-1}(-B-C)^{\lambda_2+\lambda_4-\omega-1}\\
 \notag
+& \big((\lambda_1-\lambda_4-\lambda_2+\lambda_3)A+(\lambda_1-\lambda_4+\lambda_2-\lambda_3)C-2\omega B\big)\\
 \label{3may6}
&\qquad \qquad \qquad  \qquad  \qquad  C^{\lambda_1,\lambda_4,\omega}C^{\lambda_2,\lambda_3,-\omega}
 (C-A)^{\lambda_1+\lambda_4+\omega-1}(A+C)^{\lambda_2+\lambda_3-\omega-1}\Big]=0\,.
 \end{align}}\noindent
For $C=0$ it gives 
 \begin{align}
 \label{8sep3}
     (A- \mu B) f(A,B)= k_{\Lambda-1} A^{\Lambda-1}+
     k_{\Lambda-2} A^{\Lambda-2}B+
     k_{1} AB^{\Lambda-2} +k_0 B^{\Lambda-1}\,,
 \end{align}
 where we denoted 
 \begin{align}
 \label{8sepfab}
     f(A,B)\equiv &\; \sum_{\omega} C^{\lambda_1,\lambda_2,\omega}C^{\lambda_3,\lambda_4,-\omega}
 (A-B)^{\lambda_1+\lambda_2+\omega-1}(A+B)^{\lambda_3+\lambda_4-\omega-1}\,,\\
 \notag
\mu\equiv &\; -\frac{\lambda_1-\lambda_2-\lambda_3+\lambda_4}{\lambda_1-\lambda_2+\lambda_3-\lambda_4}, \qquad 
\Lambda \equiv \lambda_1+\lambda_2+\lambda_3+\lambda_4\,,
 \end{align}
 and $k_{\Lambda-1}$, $\dots$, $k_0$  can be extracted
 from (\ref{3may6}), but will not be needed for the following 
 discussion.
 
Next we note: (i) from all possible terms of the form $A^m B^{\Lambda-m-1}$ the right hand side of (\ref{8sep3}) contains only four; (ii) $A-\mu B$ should be a divisor of the right hand side; (iii) cubic vertices with the total helicity being odd vanish, so  $f(A,B)=-f(B,A)$.
 
These three conditions allow to fix $f(A,B)$ up to an overall factor.
Indeed, the requirement that $A-\mu B$ is a divisor of the right hand side
 allows to express
 \begin{equation*}
     k_0=-k_{\Lambda-1}\mu^{\Lambda-1}-k_{\Lambda-2}\mu^{\Lambda_2}-
     k_1\mu\,.
 \end{equation*}
Then we find 
 \begin{align}
 \notag
     f(A,B)=&\; k_{\Lambda-1}A^{\Lambda-2}+(k_{\Lambda-1}\mu+k_{\Lambda-2})
     A^{\Lambda-3}B+(k_{\Lambda-1}\mu^2+k_{\Lambda-2}\mu)A^{\Lambda-4}B^2+\dots\,,\\
     \label{8sep4}
    & \qquad +(k_{\Lambda-1}\mu^{\Lambda-3}+k_{\Lambda-2}\mu^{\Lambda-4})B^{\Lambda-3}A
     +(k_1+k_{\Lambda-1}\mu^{\Lambda-2}+k_{\Lambda-2}\mu^{\Lambda-3})B^{\Lambda-2}\,.
 \end{align}
 Employing $f(A,B)=-f(B,A)$ we get, in particular,
 \begin{align*}
     k_{\Lambda-1}& = -(k_1+k_{\Lambda-1}\mu^{\Lambda-2}+k_{\Lambda-2}\mu^{\Lambda-3})\,,\\
     k_{\Lambda-1}\mu+k_{\Lambda-2}&=-(k_{\Lambda-1}\mu^{\Lambda-3}+k_{\Lambda-2}\mu^{\Lambda-4})\,.
 \end{align*}
 For real $\mu$ this implies $ k_{\Lambda-1}\mu+k_{\Lambda-2}=0$ 
 and $k_1 =-k_{\Lambda-1}$. Hence,
 \begin{equation}
 \label{8sep5}
     f(A,B)= k_{\Lambda-1}(A^{\Lambda-2}-B^{\Lambda-2}).
 \end{equation}
 Then, (\ref{8sepfab}) leads to
 \begin{equation}
 \label{9sep1}
     C^{\lambda_1,\lambda_2,\omega}C^{\lambda_3,\lambda_4,-\omega}=\frac{X(\lambda_1,\lambda_2,\lambda_3,\lambda_4)}{(\lambda_1+\lambda_2+\omega-1)!(\lambda_3+\lambda_4-\omega-1)!}\,,
 \end{equation}
 where $X(\lambda_1,\lambda_2,\lambda_3,\lambda_4)$ remains to be specified. To fix $X$,
 let us divide (\ref{9sep1}) by the same equation with $\lambda_3\to\lambda_4$ and
 $\lambda_4\to\lambda_6$
 \begin{equation*}
     \frac{C^{\lambda_3,\lambda_4,-\omega}}{C^{\lambda_5,\lambda_6,-\omega}}=
     \frac{X(\lambda_1,\lambda_2,\lambda_3,\lambda_4)}{X(\lambda_1,\lambda_2,\lambda_5,\lambda_6)}\frac{(\lambda_5+\lambda_6-\omega-1)!}{(\lambda_3+\lambda_4-\omega-1)!}\,.
 \end{equation*}
 This implies that $X$ factorises
 \begin{equation*}
     X(\lambda_1,\lambda_2,\lambda_3,\lambda_4) = Y(\lambda_1,\lambda_2)Y(\lambda_3,\lambda_4)\,.
 \end{equation*}
 Plugging this back into (\ref{9sep1}) we find 
 \begin{equation}
 \label{9sep2}
     \frac{C^{\lambda_1,\lambda_2,\omega}}{Y(\lambda_1,\lambda_2)}(\lambda_1+\lambda_2+\omega-1)!=Z(\omega)\,,
 \end{equation}
 where $Z(\omega)$ is another unknown function. Using the symmetry of $C^{\lambda_1,\lambda_2,\omega}$ we obtain
 \begin{equation*}
     Y(\lambda_{1},\lambda_2)=W Z(\lambda_1)Z(\lambda_2)
 \end{equation*}
 and hence
 \begin{equation*}
     C^{\lambda_1,\lambda_2,\omega}=W\frac{ Z(\lambda_1)Z(\lambda_2)Z(\omega)}{(\lambda_1+\lambda_2+\omega-1)!}\,.
 \end{equation*}
 Substituting this again in (\ref{9sep1}) we get
 \begin{equation*}
     Z(\omega)\cdot Z(-\omega)=1\,,
 \end{equation*}
 which is equivalent to
 \begin{equation*}
     Z(\omega)= e^{\sigma(\omega)}, \qquad \sigma(-\omega)=-\sigma(\omega)\,. 
 \end{equation*}
Eventually, we obtain that 
 \begin{equation}
 \label{9sep3}
     C^{\lambda_1,\lambda_2,\lambda_3}=W\cdot \frac{e^{\sigma(\lambda_1)+\sigma(\lambda_2)+\sigma(\lambda_3)}}{(\lambda_1+\lambda_2+\lambda_3-1)!}\,,
 \end{equation}
 where $\sigma(\lambda)$ is an arbitrary odd function.
 Substituting this into (\ref{3may6})
 we find no further constraints on $W$ and $\sigma(\lambda)$.
 So, (\ref{9sep3}) provides a general solution of the consistency condition
 (\ref{8sep1}).
 For $\sigma(\lambda)=\lambda \cdot \ln{K} $
 we reproduce the solution of Metsaev.
 Our formula (\ref{9sep3}) provides its rather obvious generalisation,
 where each spin $\lambda$ has its own coupling constant $\exp{\sigma(\lambda)}$, but they can be eaten up by rescaling  the fields. 
 
All arguments of the derivation above apply if we assume that all three spins entering each vertex  are even. Also there is a similar system for the $\bar{C}$ coefficients. 

Let us consider one more solution of the holomorphic constraints \eqref{maineq4}. Usually, the higher-spin fields are dressed by Yang-Mills groups in a stringy Chan-Paton way \cite{Konstein:1989ij,Metsaev:1991mt}. Let us instead assume that all fields take values in the adjoint of some Lie algebra of internal symmetry with structure constants $f^{a_1,a_2,a_3}$. Therefore, we should keep the space-time part of cubic vertices unchanged and multiply the coupling constants by $f^{a_1,a_2,a_3}$ 
 \begin{equation}
 \label{11sep1}
     C^{\lambda_1,\lambda_2,\lambda_3}\quad \to \quad  C^{\lambda_1,\lambda_2,\lambda_3}f^{a_1,a_2,a_3}\,.
 \end{equation}
Since the structure constants are totally anti-symmetric, this 
 changes the symmetry of the vertices to the opposite one.
Namely,  
 the vertices with the total spin being odd are totally symmetric, while
 the vertices with the total spin being even effectively vanish.
 
 Proceeding along the same lines as before we obtain a consistency
 condition which differs from (\ref{3may6}) by replacement
 (\ref{11sep1}). The consistency condition should hold
 as a consequence of the Jacobi
 identity. This implies that one has to demand
   \begin{align}
 \notag
 \sum_{\omega} \Big[& \big((\lambda_1-\lambda_2+\lambda_3-\lambda_4)A+(\lambda_1-\lambda_2-\lambda_3+\lambda_4)B-2\omega C\big)\\
 \notag
 &\qquad \qquad \qquad  \qquad  \qquad  C^{\lambda_1,\lambda_2,\omega}C^{\lambda_3,\lambda_4,-\omega}
 (A-B)^{\lambda_1+\lambda_2+\omega-1}(A+B)^{\lambda_3+\lambda_4-\omega-1}\\
 \notag
 +&\big((-\lambda_1+\lambda_3+\lambda_2-\lambda_4)B+(-\lambda_1+\lambda_3-\lambda_2+\lambda_4)C+2\omega A\big)\\
  \label{11sep2}
&\qquad \qquad \qquad  \qquad  \qquad  C^{\lambda_1,\lambda_3,\omega}C^{\lambda_2,\lambda_4,-\omega}
 (B-C)^{\lambda_1+\lambda_3+\omega-1}(-B-C)^{\lambda_2+\lambda_4-\omega-1}
 \Big]=0.
 \end{align}
 Setting $C=0$ we find that 
\begin{align}
 \label{11sep3}
     (A- \mu B) f(A,B)= 
     k_{1} AB^{\Lambda-2} +k_0 B^{\Lambda-1},
 \end{align}
 where $f(A,B)$, $\mu$ and $\Lambda$ were defined in (\ref{8sepfab}).
 By requiring that $(A-\mu B)$ is a divisor of the right hand side we
 find 
 \begin{align}
 \label{11sep4}
     f(A,B) = k_0 B^{\Lambda-2}.
 \end{align}
Due to the fact that the vertices with the total even spin vanish we
have $f(A,B) = f(B,A)$. This symmetry property is compatible with
(\ref{11sep4}) only if $\Lambda=2$. This implies 
\begin{equation}
 \label{11sep5}
     C^{\lambda_1,\lambda_2,\omega}C^{\lambda_3,\lambda_4,-\omega}=X(\lambda_1,\lambda_2,\lambda_3,\lambda_4){\delta(\lambda_1+\lambda_2+\omega-1)\delta(\lambda_3+\lambda_4-\omega-1)}.
 \end{equation}
 Proceeding as in the case of no internal symmetry,
 we obtain the solution
\begin{equation}
 \label{11sep6}
     C^{\lambda_1,\lambda_2,\lambda_3}=W\cdot {e^{\sigma(\lambda_1)+\sigma(\lambda_2)+\sigma(\lambda_3)}}{\delta(\lambda_1+\lambda_2+\lambda_3-1)}.
 \end{equation}
All arguments of the derivation above apply if we assume that all three spins entering each vertex  are odd. In the context of Chan-Paton dressing of higher-spin fields this solution was found by Metsaev in \cite{Metsaev:unpublished}.

\end{appendix}

\setstretch{1.0}
\bibliographystyle{utphys}
\bibliography{megabib.bib}

\providecommand{\href}[2]{#2}\begingroup\raggedright\begin{thebibliography}{100}

\bibitem{Weinberg:1964ew}
S.~Weinberg, ``{Photons and Gravitons in s Matrix Theory: Derivation of Charge
  Conservation and Equality of Gravitational and Inertial Mass},''
\href{http://dx.doi.org/10.1103/PhysRev.135.B1049}{{\em Phys. Rev.} {\bfseries
  135} (1964) B1049--B1056}.

\bibitem{Coleman:1967ad}
S.~R. Coleman and J.~Mandula, ``{All Possible Symmetries of the S Matrix},''
\href{http://dx.doi.org/10.1103/PhysRev.159.1251}{{\em Phys. Rev.} {\bfseries
  159} (1967) 1251--1256}.

\bibitem{Fradkin:1986qy}
E.~S. Fradkin and M.~A. Vasiliev, ``{Cubic Interaction in Extended Theories of
  Massless Higher Spin Fields},''
\href{http://dx.doi.org/10.1016/0550-3213(87)90469-X}{{\em Nucl. Phys.}
  {\bfseries B291} (1987) 141}.

\bibitem{Vasiliev:1990en}
M.~A. Vasiliev, ``Consistent equation for interacting gauge fields of all spins
  in (3+1)-dimensions,''
{\em Phys. Lett.} {\bfseries B243} (1990) 378--382.

\bibitem{Vasiliev:2003ev}
M.~A. Vasiliev, ``{Nonlinear equations for symmetric massless higher spin
  fields in (A)dS(d)},''
  \href{http://dx.doi.org/10.1016/S0370-2693(03)00872-4}{{\em Phys. Lett.}
  {\bfseries B567} (2003) 139--151},
\href{http://arxiv.org/abs/hep-th/0304049}{{\ttfamily arXiv:hep-th/0304049
  [hep-th]}}.

\bibitem{Maldacena:1997re}
J.~M. Maldacena, ``{The large N limit of superconformal field theories and
  supergravity},'' {\em Adv. Theor. Math. Phys.} {\bfseries 2} (1998) 231--252,
\href{http://arxiv.org/abs/hep-th/9711200}{{\ttfamily arXiv:hep-th/9711200}}.

\bibitem{Gubser:1998bc}
S.~S. Gubser, I.~R. Klebanov, and A.~M. Polyakov, ``{Gauge theory correlators
  from non-critical string theory},''
  \href{http://dx.doi.org/10.1016/S0370-2693(98)00377-3}{{\em Phys. Lett.}
  {\bfseries B428} (1998) 105--114},
\href{http://arxiv.org/abs/hep-th/9802109}{{\ttfamily arXiv:hep-th/9802109}}.

\bibitem{Witten:1998qj}
E.~Witten, ``{Anti-de Sitter space and holography},'' {\em Adv. Theor. Math.
  Phys.} {\bfseries 2} (1998) 253--291,
\href{http://arxiv.org/abs/hep-th/9802150}{{\ttfamily arXiv:hep-th/9802150}}.

\bibitem{Klebanov:2002ja}
I.~R. Klebanov and A.~M. Polyakov, ``{AdS dual of the critical O(N) vector
  model},'' \href{http://dx.doi.org/10.1016/S0370-2693(02)02980-5}{{\em Phys.
  Lett.} {\bfseries B550} (2002) 213--219},
\href{http://arxiv.org/abs/hep-th/0210114}{{\ttfamily arXiv:hep-th/0210114}}.

\bibitem{Sezgin:2002rt}
E.~Sezgin and P.~Sundell, ``{Massless higher spins and holography},''
  \href{http://dx.doi.org/10.1016/S0550-3213(02)00739-3}{{\em Nucl.Phys.}
  {\bfseries B644} (2002) 303--370},
\href{http://arxiv.org/abs/hep-th/0205131}{{\ttfamily arXiv:hep-th/0205131
  [hep-th]}}.

\bibitem{Sezgin:2003pt}
E.~Sezgin and P.~Sundell, ``{Holography in 4D (super) higher spin theories and
  a test via cubic scalar couplings},''
  \href{http://dx.doi.org/10.1088/1126-6708/2005/07/044}{{\em JHEP} {\bfseries
  0507} (2005) 044},
\href{http://arxiv.org/abs/hep-th/0305040}{{\ttfamily arXiv:hep-th/0305040
  [hep-th]}}.

\bibitem{Giombi:2009wh}
S.~Giombi and X.~Yin, ``{Higher Spin Gauge Theory and Holography: The
  Three-Point Functions},''
  \href{http://dx.doi.org/10.1007/JHEP09(2010)115}{{\em JHEP} {\bfseries 1009}
  (2010) 115},
\href{http://arxiv.org/abs/0912.3462}{{\ttfamily arXiv:0912.3462 [hep-th]}}.

\bibitem{Klebanov:1999tb}
I.~R. Klebanov and E.~Witten, ``{AdS / CFT correspondence and symmetry
  breaking},'' \href{http://dx.doi.org/10.1016/S0550-3213(99)00387-9}{{\em
  Nucl. Phys.} {\bfseries B556} (1999) 89--114},
\href{http://arxiv.org/abs/hep-th/9905104}{{\ttfamily arXiv:hep-th/9905104
  [hep-th]}}.

\bibitem{Leigh:2003gk}
R.~G. Leigh and A.~C. Petkou, ``{Holography of the N=1 higher spin theory on
  AdS(4)},'' {\em JHEP} {\bfseries 0306} (2003) 011,
\href{http://arxiv.org/abs/hep-th/0304217}{{\ttfamily arXiv:hep-th/0304217
  [hep-th]}}.

\bibitem{Giombi:2011ya}
S.~Giombi and X.~Yin, ``{On Higher Spin Gauge Theory and the Critical O(N)
  Model},'' \href{http://dx.doi.org/10.1103/PhysRevD.85.086005}{{\em Phys.Rev.}
  {\bfseries D85} (2012) 086005},
\href{http://arxiv.org/abs/1105.4011}{{\ttfamily arXiv:1105.4011 [hep-th]}}.

\bibitem{Aragone:1979hx}
C.~Aragone and S.~Deser, ``{Consistency Problems of Hypergravity},''
\href{http://dx.doi.org/10.1016/0370-2693(79)90808-6}{{\em Phys. Lett.}
  {\bfseries B86} (1979) 161--163}.

\bibitem{Aragone:1981yn}
C.~Aragone and H.~La~Roche, ``{Massless Second Order Tetradic Spin 3 Fields and
  Higher Helicity Bosons},''
\href{http://dx.doi.org/10.1007/BF02902412}{{\em Nuovo Cim.} {\bfseries A72}
  (1982) 149}.

\bibitem{Bekaert:2010hp}
X.~Bekaert, N.~Boulanger, and S.~Leclercq, ``{Strong obstruction of the
  Berends-Burgers-van Dam spin-3 vertex},''
  \href{http://dx.doi.org/10.1088/1751-8113/43/18/185401}{{\em J. Phys.}
  {\bfseries A43} (2010) 185401},
\href{http://arxiv.org/abs/1002.0289}{{\ttfamily arXiv:1002.0289 [hep-th]}}.

\bibitem{Joung:2013nma}
E.~Joung and M.~Taronna, ``{Cubic-interaction-induced deformations of
  higher-spin symmetries},''
  \href{http://dx.doi.org/10.1007/JHEP03(2014)103}{{\em JHEP} {\bfseries 03}
  (2014) 103},
\href{http://arxiv.org/abs/1311.0242}{{\ttfamily arXiv:1311.0242 [hep-th]}}.

\bibitem{Fronsdal:1978rb}
C.~Fronsdal, ``Massless fields with integer spin,''
{\em Phys. Rev.} {\bfseries D18} (1978) 3624.

\bibitem{Bengtsson:1983pg}
A.~K.~H. Bengtsson, I.~Bengtsson, and L.~Brink, ``{Cubic Interaction Terms for
  Arbitrarily Extended Supermultiplets},''
\href{http://dx.doi.org/10.1016/0550-3213(83)90141-4}{{\em Nucl. Phys.}
  {\bfseries B227} (1983) 41--49}.

\bibitem{Bengtsson:1983pd}
A.~K.~H. Bengtsson, I.~Bengtsson, and L.~Brink, ``{Cubic Interaction Terms for
  Arbitrary Spin},''
\href{http://dx.doi.org/10.1016/0550-3213(83)90140-2}{{\em Nucl. Phys.}
  {\bfseries B227} (1983) 31--40}.

\bibitem{Bengtsson:1986kh}
A.~K.~H. Bengtsson, I.~Bengtsson, and N.~Linden, ``{Interacting Higher Spin
  Gauge Fields on the Light Front},''
\href{http://dx.doi.org/10.1088/0264-9381/4/5/028}{{\em Class. Quant. Grav.}
  {\bfseries 4} (1987) 1333}.

\bibitem{Berends:1984wp}
F.~A. Berends, G.~J.~H. Burgers, and H.~Van~Dam, ``{On spin three
  selfinteractions},''
\href{http://dx.doi.org/10.1007/BF01410362}{{\em Z. Phys.} {\bfseries C24}
  (1984) 247--254}.

\bibitem{Berends:1984rq}
F.~A. Berends, G.~J.~H. Burgers, and H.~van Dam, ``{On the Theoretical Problems
  in Constructing Interactions Involving Higher Spin Massless Particles},''
\href{http://dx.doi.org/10.1016/0550-3213(85)90074-4}{{\em Nucl. Phys.}
  {\bfseries B260} (1985) 295--322}.

\bibitem{Fradkin:1991iy}
E.~S. Fradkin and R.~R. Metsaev, ``{A Cubic interaction of totally symmetric
  massless representations of the Lorentz group in arbitrary dimensions},''
\href{http://dx.doi.org/10.1088/0264-9381/8/4/004}{{\em Class. Quant. Grav.}
  {\bfseries 8} (1991) L89--L94}.

\bibitem{Metsaev:1993ap}
R.~R. Metsaev, ``{Generating function for cubic interaction vertices of higher
  spin fields in any dimension},''
\href{http://dx.doi.org/10.1142/S0217732393003706}{{\em Mod. Phys. Lett.}
  {\bfseries A8} (1993) 2413--2426}.

\bibitem{Metsaev:2005ar}
R.~R. Metsaev, ``{Cubic interaction vertices of massive and massless higher
  spin fields},'' \href{http://dx.doi.org/10.1016/j.nuclphysb.2006.10.002}{{\em
  Nucl. Phys.} {\bfseries B759} (2006) 147--201},
\href{http://arxiv.org/abs/hep-th/0512342}{{\ttfamily arXiv:hep-th/0512342
  [hep-th]}}.

\bibitem{Metsaev:2007rn}
R.~R. Metsaev, ``{Cubic interaction vertices for fermionic and bosonic
  arbitrary spin fields},''
  \href{http://dx.doi.org/10.1016/j.nuclphysb.2012.01.022}{{\em Nucl. Phys.}
  {\bfseries B859} (2012) 13--69},
\href{http://arxiv.org/abs/0712.3526}{{\ttfamily arXiv:0712.3526 [hep-th]}}.

\bibitem{Bengtsson:2014qza}
A.~K.~H. Bengtsson, ``{A Riccati type PDE for light-front higher helicity
  vertices},'' \href{http://dx.doi.org/10.1007/JHEP09(2014)105}{{\em JHEP}
  {\bfseries 09} (2014) 105},
\href{http://arxiv.org/abs/1403.7345}{{\ttfamily arXiv:1403.7345 [hep-th]}}.

\bibitem{Conde:2016izb}
E.~Conde, E.~Joung, and K.~Mkrtchyan, ``{Spinor-Helicity Three-Point Amplitudes
  from Local Cubic Interactions},''
  \href{http://dx.doi.org/10.1007/JHEP08(2016)040}{{\em JHEP} {\bfseries 08}
  (2016) 040},
\href{http://arxiv.org/abs/1605.07402}{{\ttfamily arXiv:1605.07402 [hep-th]}}.

\bibitem{Sleight:2016xqq}
C.~Sleight and M.~Taronna, ``{Higher-Spin Algebras, Holography and Flat
  Space},''
\href{http://arxiv.org/abs/1609.00991}{{\ttfamily arXiv:1609.00991 [hep-th]}}.

\bibitem{Benincasa:2011pg}
P.~Benincasa and E.~Conde, ``{Exploring the S-Matrix of Massless Particles},''
  \href{http://dx.doi.org/10.1103/PhysRevD.86.025007}{{\em Phys. Rev.}
  {\bfseries D86} (2012) 025007},
\href{http://arxiv.org/abs/1108.3078}{{\ttfamily arXiv:1108.3078 [hep-th]}}.

\bibitem{Benincasa:2007xk}
P.~Benincasa and F.~Cachazo, ``{Consistency Conditions on the S-Matrix of
  Massless Particles},''
\href{http://arxiv.org/abs/0705.4305}{{\ttfamily arXiv:0705.4305 [hep-th]}}.

\bibitem{Metsaev:1991nb}
R.~R. Metsaev, ``{S matrix approach to massless higher spins theory. 2: The
  Case of internal symmetry},''
\href{http://dx.doi.org/10.1142/S0217732391002839}{{\em Mod. Phys. Lett.}
  {\bfseries A6} (1991) 2411--2421}.

\bibitem{Metsaev:1991mt}
R.~R. Metsaev, ``{Poincare invariant dynamics of massless higher spins: Fourth
  order analysis on mass shell},''
\href{http://dx.doi.org/10.1142/S0217732391000348}{{\em Mod. Phys. Lett.}
  {\bfseries A6} (1991) 359--367}.

\bibitem{Bekaert:2014cea}
X.~Bekaert, J.~Erdmenger, D.~Ponomarev, and C.~Sleight, ``{Towards holographic
  higher-spin interactions: Four-point functions and higher-spin exchange},''
  \href{http://dx.doi.org/10.1007/JHEP03(2015)170}{{\em JHEP} {\bfseries 03}
  (2015) 170},
\href{http://arxiv.org/abs/1412.0016}{{\ttfamily arXiv:1412.0016 [hep-th]}}.

\bibitem{Bekaert:2015tva}
X.~Bekaert, J.~Erdmenger, D.~Ponomarev, and C.~Sleight, ``{Quartic AdS
  Interactions in Higher-Spin Gravity from Conformal Field Theory},''
  \href{http://dx.doi.org/10.1007/JHEP11(2015)149}{{\em JHEP} {\bfseries 11}
  (2015) 149},
\href{http://arxiv.org/abs/1508.04292}{{\ttfamily arXiv:1508.04292 [hep-th]}}.

\bibitem{Bekaert:2010hw}
X.~Bekaert, N.~Boulanger, and P.~Sundell, ``{How higher-spin gravity surpasses
  the spin two barrier: no-go theorems versus yes-go examples},''
  \href{http://dx.doi.org/10.1103/RevModPhys.84.987}{{\em Rev.Mod.Phys.}
  {\bfseries 84} (2012) 987--1009},
\href{http://arxiv.org/abs/1007.0435}{{\ttfamily arXiv:1007.0435 [hep-th]}}.

\bibitem{Beccaria:2016syk}
M.~Beccaria, S.~Nakach, and A.~A. Tseytlin, ``{On triviality of S-matrix in
  conformal higher spin theory},''
\href{http://arxiv.org/abs/1607.06379}{{\ttfamily arXiv:1607.06379 [hep-th]}}.

\bibitem{Joung:2015eny}
E.~Joung, S.~Nakach, and A.~A. Tseytlin, ``{Scalar scattering via conformal
  higher spin exchange},''
  \href{http://dx.doi.org/10.1007/JHEP02(2016)125}{{\em JHEP} {\bfseries 02}
  (2016) 125},
\href{http://arxiv.org/abs/1512.08896}{{\ttfamily arXiv:1512.08896 [hep-th]}}.

\bibitem{Maldacena:2011jn}
J.~Maldacena and A.~Zhiboedov, ``{Constraining Conformal Field Theories with A
  Higher Spin Symmetry},''
\href{http://arxiv.org/abs/1112.1016}{{\ttfamily arXiv:1112.1016 [hep-th]}}.

\bibitem{Alba:2013yda}
V.~Alba and K.~Diab, ``{Constraining conformal field theories with a higher
  spin symmetry in d=4},''
\href{http://arxiv.org/abs/1307.8092}{{\ttfamily arXiv:1307.8092 [hep-th]}}.

\bibitem{Boulanger:2013zza}
N.~Boulanger, D.~Ponomarev, E.~Skvortsov, and M.~Taronna, ``{On the uniqueness
  of higher-spin symmetries in AdS and CFT},''
\href{http://arxiv.org/abs/1305.5180}{{\ttfamily arXiv:1305.5180 [hep-th]}}.

\bibitem{Stanev:2013qra}
Y.~S. Stanev, ``{Constraining conformal field theory with higher spin symmetry
  in four dimensions},''
  \href{http://dx.doi.org/10.1016/j.nuclphysb.2013.09.002}{{\em Nucl. Phys.}
  {\bfseries B876} (2013) 651--666},
\href{http://arxiv.org/abs/1307.5209}{{\ttfamily arXiv:1307.5209 [hep-th]}}.

\bibitem{Alba:2015upa}
V.~Alba and K.~Diab, ``{Constraining conformal field theories with a higher
  spin symmetry in $d> 3$ dimensions},''
\href{http://arxiv.org/abs/1510.02535}{{\ttfamily arXiv:1510.02535 [hep-th]}}.

\bibitem{Fronsdal:1978vb}
C.~Fronsdal, ``{Singletons and Massless, Integral Spin Fields on de Sitter
  Space (Elementary Particles in a Curved Space. 7.},''
\href{http://dx.doi.org/10.1103/PhysRevD.20.848}{{\em Phys.Rev.} {\bfseries
  D20} (1979) 848--856}.

\bibitem{Fradkin:1986ka}
E.~S. Fradkin and M.~A. Vasiliev, ``Candidate to the role of higher spin
  symmetry,''
{\em Ann. Phys.} {\bfseries 177} (1987) 63.

\bibitem{commMassimo}
 Using the classification of cubic couplings the assumption of finite number of
  fields was removed in \cite{Joung:2013nma}. This approach is different from
  that of the proof of the Coleman-Mandula theorem.

\bibitem{Boulanger:2006gr}
N.~Boulanger and S.~Leclercq, ``{Consistent couplings between spin-2 and spin-3
  massless fields},'' {\em JHEP} {\bfseries 11} (2006) 034,
\href{http://arxiv.org/abs/hep-th/0609221}{{\ttfamily arXiv:hep-th/0609221}}.

\bibitem{Zinoviev:2008ck}
{\relax Yu}.~M. Zinoviev, ``{On spin 3 interacting with gravity},''
  \href{http://dx.doi.org/10.1088/0264-9381/26/3/035022}{{\em Class. Quant.
  Grav.} {\bfseries 26} (2009) 035022},
\href{http://arxiv.org/abs/0805.2226}{{\ttfamily arXiv:0805.2226 [hep-th]}}.

\bibitem{Boulanger:2008tg}
N.~Boulanger, S.~Leclercq, and P.~Sundell, ``{On The Uniqueness of Minimal
  Coupling in Higher-Spin Gauge Theory},''
  \href{http://dx.doi.org/10.1088/1126-6708/2008/08/056}{{\em JHEP} {\bfseries
  08} (2008) 056},
\href{http://arxiv.org/abs/0805.2764}{{\ttfamily arXiv:0805.2764 [hep-th]}}.

\bibitem{Fotopoulos:2010ay}
A.~Fotopoulos and M.~Tsulaia, ``{On the Tensionless Limit of String theory, Off
  - Shell Higher Spin Interaction Vertices and BCFW Recursion Relations},''
  \href{http://dx.doi.org/10.1007/JHEP11(2010)086}{{\em JHEP} {\bfseries 11}
  (2010) 086},
\href{http://arxiv.org/abs/1009.0727}{{\ttfamily arXiv:1009.0727 [hep-th]}}.

\bibitem{McGady:2013sga}
D.~A. McGady and L.~Rodina, ``{Higher-spin massless $S$-matrices in
  four-dimensions},'' \href{http://dx.doi.org/10.1103/PhysRevD.90.084048}{{\em
  Phys. Rev.} {\bfseries D90} no.~8, (2014) 084048},
\href{http://arxiv.org/abs/1311.2938}{{\ttfamily arXiv:1311.2938 [hep-th]}}.

\bibitem{Bengtsson:2016alt}
A.~K.~H. Bengtsson, ``{Quartic amplitudes for Minkowski higher spin},'' in {\em
  {International Workshop on Higher Spin Gauge Theories Singapore, Singapore,
  November 4-6, 2015}}.
\newblock 2016.
\newblock
\href{http://arxiv.org/abs/1605.02608}{{\ttfamily arXiv:1605.02608 [hep-th]}}.
\newblock

\bibitem{Ponomarev:2016jqk}
D.~Ponomarev and A.~A. Tseytlin, ``{On quantum corrections in higher-spin
  theory in flat space},''
  \href{http://dx.doi.org/10.1007/JHEP05(2016)184}{{\em JHEP} {\bfseries 05}
  (2016) 184},
\href{http://arxiv.org/abs/1603.06273}{{\ttfamily arXiv:1603.06273 [hep-th]}}.

\bibitem{Blencowe:1988gj}
M.~Blencowe, ``{A Consistent Interacting Massless Higher Spin Field Theory in
  $D$ = (2+1)},''
\href{http://dx.doi.org/10.1088/0264-9381/6/4/005}{{\em Class.Quant.Grav.}
  {\bfseries 6} (1989) 443}.

\bibitem{Campoleoni:2010zq}
A.~Campoleoni, S.~Fredenhagen, S.~Pfenninger, and S.~Theisen, ``{Asymptotic
  symmetries of three-dimensional gravity coupled to higher-spin fields},''
  \href{http://dx.doi.org/10.1007/JHEP11(2010)007}{{\em JHEP} {\bfseries 1011}
  (2010) 007},
\href{http://arxiv.org/abs/1008.4744}{{\ttfamily arXiv:1008.4744 [hep-th]}}.

\bibitem{Henneaux:2010xg}
M.~Henneaux and S.-J. Rey, ``{Nonlinear $W_{infinity}$ as Asymptotic Symmetry
  of Three-Dimensional Higher Spin Anti-de Sitter Gravity},''
  \href{http://dx.doi.org/10.1007/JHEP12(2010)007}{{\em JHEP} {\bfseries 12}
  (2010) 007},
\href{http://arxiv.org/abs/1008.4579}{{\ttfamily arXiv:1008.4579 [hep-th]}}.

\bibitem{Gaberdiel:2012uj}
M.~R. Gaberdiel and R.~Gopakumar, ``{Minimal Model Holography},''
  \href{http://dx.doi.org/10.1088/1751-8113/46/21/214002}{{\em J. Phys.}
  {\bfseries A46} (2013) 214002},
\href{http://arxiv.org/abs/1207.6697}{{\ttfamily arXiv:1207.6697 [hep-th]}}.

\bibitem{Afshar:2013vka}
H.~Afshar, A.~Bagchi, R.~Fareghbal, D.~Grumiller, and J.~Rosseel, ``{Spin-3
  Gravity in Three-Dimensional Flat Space},''
  \href{http://dx.doi.org/10.1103/PhysRevLett.111.121603}{{\em Phys. Rev.
  Lett.} {\bfseries 111} no.~12, (2013) 121603},
\href{http://arxiv.org/abs/1307.4768}{{\ttfamily arXiv:1307.4768 [hep-th]}}.

\bibitem{Gonzalez:2013oaa}
H.~A. Gonzalez, J.~Matulich, M.~Pino, and R.~Troncoso, ``{Asymptotically flat
  spacetimes in three-dimensional higher spin gravity},''
  \href{http://dx.doi.org/10.1007/JHEP09(2013)016}{{\em JHEP} {\bfseries 09}
  (2013) 016},
\href{http://arxiv.org/abs/1307.5651}{{\ttfamily arXiv:1307.5651 [hep-th]}}.

\bibitem{Bengtsson:2012jm}
A.~K.~H. Bengtsson, ``{Systematics of Higher-spin Light-front Interactions},''
\newblock 2012.
\newblock
\href{http://arxiv.org/abs/1205.6117}{{\ttfamily arXiv:1205.6117 [hep-th]}}.
\newblock

\bibitem{Dirac:1949cp}
P.~A.~M. Dirac, ``{Forms of Relativistic Dynamics},''
\href{http://dx.doi.org/10.1103/RevModPhys.21.392}{{\em Rev. Mod. Phys.}
  {\bfseries 21} (1949) 392--399}.

\bibitem{Gitman:1990qh}
D.~M. Gitman and I.~V. Tyutin, {\em {Quantization of Fields with Constraints}}.
\newblock Springer Series in Nuclear and Particle Physics. Springer, Berlin,
  Germany,
1990.
\newblock

\bibitem{Henneaux:1992ig}
M.~Henneaux and C.~Teitelboim, {\em {Quantization of gauge systems}}.
\newblock
1992.
\newblock

\bibitem{Ananth:2012un}
S.~Ananth, ``{Spinor helicity structures in higher spin theories},''
  \href{http://dx.doi.org/10.1007/JHEP11(2012)089}{{\em JHEP} {\bfseries 11}
  (2012) 089},
\href{http://arxiv.org/abs/1209.4960}{{\ttfamily arXiv:1209.4960 [hep-th]}}.

\bibitem{Akshay:2014qea}
Y.~S. Akshay and S.~Ananth, ``{Factorization of cubic vertices involving three
  different higher spin fields},''
  \href{http://dx.doi.org/10.1016/j.nuclphysb.2014.08.002}{{\em Nucl. Phys.}
  {\bfseries B887} (2014) 168--174},
\href{http://arxiv.org/abs/1404.2448}{{\ttfamily arXiv:1404.2448 [hep-th]}}.

\bibitem{Bengtsson:2016jfk}
A.~K.~H. Bengtsson, ``{Notes on Cubic and Quartic Light-Front Kinematics},''
\href{http://arxiv.org/abs/1604.01974}{{\ttfamily arXiv:1604.01974
  [physics.gen-ph]}}.

\bibitem{Buchbinder:2006eq}
I.~L. Buchbinder, A.~Fotopoulos, A.~C. Petkou, and M.~Tsulaia, ``{Constructing
  the cubic interaction vertex of higher spin gauge fields},''
  \href{http://dx.doi.org/10.1103/PhysRevD.74.105018}{{\em Phys. Rev.}
  {\bfseries D74} (2006) 105018},
\href{http://arxiv.org/abs/hep-th/0609082}{{\ttfamily arXiv:hep-th/0609082
  [hep-th]}}.

\bibitem{Manvelyan:2010je}
R.~Manvelyan, K.~Mkrtchyan, and W.~Ruehl, ``{A Generating function for the
  cubic interactions of higher spin fields},''
  \href{http://dx.doi.org/10.1016/j.physletb.2010.12.049}{{\em Phys. Lett.}
  {\bfseries B696} (2011) 410--415},
\href{http://arxiv.org/abs/1009.1054}{{\ttfamily arXiv:1009.1054 [hep-th]}}.

\bibitem{Zinoviev:2010cr}
{\relax Yu}.~M. Zinoviev, ``{Spin 3 cubic vertices in a frame-like
  formalism},'' \href{http://dx.doi.org/10.1007/JHEP08(2010)084}{{\em JHEP}
  {\bfseries 08} (2010) 084},
\href{http://arxiv.org/abs/1007.0158}{{\ttfamily arXiv:1007.0158 [hep-th]}}.

\bibitem{Taronna:2011kt}
M.~Taronna, ``{Higher-Spin Interactions: four-point functions and beyond},''
  \href{http://dx.doi.org/10.1007/JHEP04(2012)029}{{\em JHEP} {\bfseries 04}
  (2012) 029},
\href{http://arxiv.org/abs/1107.5843}{{\ttfamily arXiv:1107.5843 [hep-th]}}.

\bibitem{Berkovits:1997wj}
N.~Berkovits and C.~M. Hull, ``{Manifestly covariant actions for D = 4 selfdual
  Yang-Mills and D = 10 superYang-Mills},''
  \href{http://dx.doi.org/10.1088/1126-6708/1998/02/012}{{\em JHEP} {\bfseries
  02} (1998) 012},
\href{http://arxiv.org/abs/hep-th/9712007}{{\ttfamily arXiv:hep-th/9712007
  [hep-th]}}.

\bibitem{Pasti:2009xc}
P.~Pasti, I.~Samsonov, D.~Sorokin, and M.~Tonin, ``{BLG-motivated Lagrangian
  formulation for the chiral two-form gauge field in D=6 and M5-branes},''
  \href{http://dx.doi.org/10.1103/PhysRevD.80.086008}{{\em Phys. Rev.}
  {\bfseries D80} (2009) 086008},
\href{http://arxiv.org/abs/0907.4596}{{\ttfamily arXiv:0907.4596 [hep-th]}}.

\bibitem{Henneaux:1988gg}
M.~Henneaux and C.~Teitelboim, ``{Dynamics of Chiral (Selfdual) $P$ Forms},''
\href{http://dx.doi.org/10.1016/0370-2693(88)90712-5}{{\em Phys. Lett.}
  {\bfseries B206} (1988) 650--654}.

\bibitem{Giombi:2011kc}
S.~Giombi, S.~Minwalla, S.~Prakash, S.~P. Trivedi, S.~R. Wadia, and X.~Yin,
  ``{Chern-Simons Theory with Vector Fermion Matter},''
  \href{http://dx.doi.org/10.1140/epjc/s10052-012-2112-0}{{\em Eur. Phys. J.}
  {\bfseries C72} (2012) 2112},
\href{http://arxiv.org/abs/1110.4386}{{\ttfamily arXiv:1110.4386 [hep-th]}}.

\bibitem{Barnich:1993vg}
G.~Barnich and M.~Henneaux, ``{Consistent couplings between fields with a gauge
  freedom and deformations of the master equation},''
  \href{http://dx.doi.org/10.1016/0370-2693(93)90544-R}{{\em Phys. Lett.}
  {\bfseries B311} (1993) 123--129},
\href{http://arxiv.org/abs/hep-th/9304057}{{\ttfamily arXiv:hep-th/9304057
  [hep-th]}}.

\bibitem{Metsaev:unpublished}
R.~R. Metsaev, ``{Ph.D. Thesis}.'' 1991.

\bibitem{Skvortsov:2015pea}
E.~D. Skvortsov, ``{On (Un)Broken Higher-Spin Symmetry in Vector Models},''
\href{http://arxiv.org/abs/hep-th:1512.05994}{{\ttfamily
  arXiv:hep-th:1512.05994 [hep-th]}}.

\bibitem{Sleight:2016dba}
C.~Sleight and M.~Taronna, ``{Higher Spin Interactions from Conformal Field
  Theory: The Complete Cubic Couplings},''
  \href{http://dx.doi.org/10.1103/PhysRevLett.116.181602}{{\em Phys. Rev.
  Lett.} {\bfseries 116} no.~18, (2016) 181602},
\href{http://arxiv.org/abs/1603.00022}{{\ttfamily arXiv:1603.00022 [hep-th]}}.

\bibitem{Haehnel:2016mlb}
P.~Haehnel and T.~McLoughlin, ``{Conformal Higher Spin Theory and Twistor Space
  Actions},''
\href{http://arxiv.org/abs/1604.08209}{{\ttfamily arXiv:1604.08209 [hep-th]}}.

\bibitem{Giombi:2013yva}
S.~Giombi, I.~R. Klebanov, S.~S. Pufu, B.~R. Safdi, and G.~Tarnopolsky, ``{AdS
  Description of Induced Higher-Spin Gauge Theory},''
  \href{http://dx.doi.org/10.1007/JHEP10(2013)016}{{\em JHEP} {\bfseries 10}
  (2013) 016},
\href{http://arxiv.org/abs/1306.5242}{{\ttfamily arXiv:1306.5242 [hep-th]}}.

\bibitem{Soojong}
S.-J. Rey  (to appear) .

\bibitem{Vasiliev:1989re}
M.~A. Vasiliev, ``Higher spin algebras and quantization on the sphere and
  hyperboloid,''
{\em Int. J. Mod. Phys.} {\bfseries A6} (1991) 1115--1135.

\bibitem{Feigin}
B.~Feigin, ``{The Lie algebras gl(l) and cohomologies of Lie algebras of
  differential operators},'' \href{http://dx.doi.org/0036-0279/43/2/L12}{{\em
  Russ. Math. Surv.} {\bfseries 34} (1988) 169}.

\bibitem{Gunaydin:1984fk}
M.~Gunaydin and N.~Marcus, ``{The Spectrum of the $S^5$ Compactification of the
  Chiral N=2, D=10 Supergravity and the Unitary Supermultiplets of $U(2,
  2/4)$},''
\href{http://dx.doi.org/10.1088/0264-9381/2/2/001}{{\em Class. Quant. Grav.}
  {\bfseries 2} (1985) L11}.

\bibitem{Gunaydin:1984wc}
M.~Gunaydin, P.~van Nieuwenhuizen, and N.~P. Warner, ``{General Construction of
  the Unitary Representations of Anti-de Sitter Superalgebras and the Spectrum
  of the S(4) Compactification of Eleven-dimensional Supergravity},''
\href{http://dx.doi.org/10.1016/0550-3213(85)90129-4}{{\em Nucl. Phys.}
  {\bfseries B255} (1985) 63--92}.

\bibitem{Gunaydin:2016amv}
M.~Gunaydin, E.~D. Skvortsov, and T.~Tran, ``{Exceptional F(4) Higher-Spin
  Theory in AdS(6) at One-Loop and other Tests of Duality},''
\href{http://arxiv.org/abs/1608.07582}{{\ttfamily arXiv:1608.07582 [hep-th]}}.

\bibitem{Giombi:2013fka}
S.~Giombi and I.~R. Klebanov, ``{One Loop Tests of Higher Spin AdS/CFT},''
  \href{http://dx.doi.org/10.1007/JHEP12(2013)068}{{\em JHEP} {\bfseries 12}
  (2013) 068},
\href{http://arxiv.org/abs/1308.2337}{{\ttfamily arXiv:1308.2337 [hep-th]}}.

\bibitem{Giombi:2014iua}
S.~Giombi, I.~R. Klebanov, and B.~R. Safdi, ``{Higher Spin $AdS_{d+1}/CFT_d$ at
  One Loop},'' \href{http://dx.doi.org/10.1103/PhysRevD.89.084004}{{\em Phys.
  Rev.} {\bfseries D89} no.~8, (2014) 084004},
\href{http://arxiv.org/abs/1401.0825}{{\ttfamily arXiv:1401.0825 [hep-th]}}.

\bibitem{Giombi:2014yra}
S.~Giombi, I.~R. Klebanov, and A.~A. Tseytlin, ``{Partition Functions and
  Casimir Energies in Higher Spin $AdS_{d+1}/CFT_d$},''
  \href{http://dx.doi.org/10.1103/PhysRevD.90.024048}{{\em Phys. Rev.}
  {\bfseries D90} no.~2, (2014) 024048},
\href{http://arxiv.org/abs/1402.5396}{{\ttfamily arXiv:1402.5396 [hep-th]}}.

\bibitem{Fradkin:1989md}
E.~S. Fradkin and V.~{\relax Ya}. Linetsky, ``{Cubic Interaction in Conformal
  Theory of Integer Higher Spin Fields in Four-dimensional Space-time},''
\href{http://dx.doi.org/10.1016/0370-2693(89)90120-2}{{\em Phys. Lett.}
  {\bfseries B231} (1989) 97--106}.

\bibitem{Fernando:2009fq}
S.~Fernando and M.~Gunaydin, ``{Minimal unitary representation of SU(2,2) and
  its deformations as massless conformal fields and their supersymmetric
  extensions},'' \href{http://dx.doi.org/10.1063/1.3447773}{{\em J.Math.Phys.}
  {\bfseries 51} (2010) 082301},
\href{http://arxiv.org/abs/0908.3624}{{\ttfamily arXiv:0908.3624 [hep-th]}}.

\bibitem{Boulanger:2011se}
N.~Boulanger and E.~Skvortsov, ``{Higher-spin algebras and cubic interactions
  for simple mixed-symmetry fields in AdS spacetime},''
  \href{http://dx.doi.org/10.1007/JHEP09(2011)063}{{\em JHEP} {\bfseries 1109}
  (2011) 063},
\href{http://arxiv.org/abs/1107.5028}{{\ttfamily arXiv:1107.5028 [hep-th]}}.

\bibitem{Govil:2013uta}
K.~Govil and M.~Gunaydin, ``{Deformed Twistors and Higher Spin Conformal
  (Super-)Algebras in Four Dimensions},''
  \href{http://dx.doi.org/10.1007/JHEP07(2014)004}{{\em JHEP} {\bfseries 03}
  (2015) 026},
\href{http://arxiv.org/abs/1312.2907}{{\ttfamily arXiv:1312.2907 [hep-th]}}.

\bibitem{Manvelyan:2013oua}
R.~Manvelyan, K.~Mkrtchyan, R.~Mkrtchyan, and S.~Theisen, ``{On Higher Spin
  Symmetries in $AdS_{5}$},''
  \href{http://dx.doi.org/10.1007/JHEP10(2013)185}{{\em JHEP} {\bfseries 10}
  (2013) 185},
\href{http://arxiv.org/abs/1304.7988}{{\ttfamily arXiv:1304.7988 [hep-th]}}.

\bibitem{Joung:2014qya}
E.~Joung and K.~Mkrtchyan, ``{Notes on higher-spin algebras: minimal
  representations and structure constants},''
  \href{http://dx.doi.org/10.1007/JHEP05(2014)103}{{\em JHEP} {\bfseries 05}
  (2014) 103},
\href{http://arxiv.org/abs/1401.7977}{{\ttfamily arXiv:1401.7977 [hep-th]}}.

\bibitem{Caron-Huot:2016icg}
S.~Caron-Huot, Z.~Komargodski, A.~Sever, and A.~Zhiboedov, ``{Strings from
  Massive Higher Spins: The Asymptotic Uniqueness of the Veneziano
  Amplitude},''
\href{http://arxiv.org/abs/1607.04253}{{\ttfamily arXiv:1607.04253 [hep-th]}}.

\bibitem{Konstein:1989ij}
S.~E. Konstein and M.~A. Vasiliev, ``Extended higher spin superalgebras and
  their massless representations,''
{\em Nucl. Phys.} {\bfseries B331} (1990) 475--499.

\end{thebibliography}\endgroup
\end{document}